\documentclass[conference]{IEEETran}
\IEEEoverridecommandlockouts
\usepackage{cite}
\usepackage{amsmath,amssymb,amsfonts}
\usepackage{graphicx}
\usepackage{textcomp}
\usepackage{xcolor}
\usepackage{booktabs} 
\usepackage{bm}
\usepackage{amsmath}
\usepackage{amssymb}
\usepackage{setspace}
\usepackage{cases}
\usepackage{enumitem}
\usepackage{varwidth}
\usepackage{makecell}
\usepackage{url}
\usepackage{comment}
\usepackage{array}
\usepackage[ruled]{algorithm2e} 
\usepackage{graphicx}
\usepackage{multirow}
\usepackage{lipsum}
\usepackage{capt-of}
\usepackage{epstopdf}
\usepackage{comment}
\usepackage{authblk}
\usepackage{color, colortbl}
\usepackage[normalem]{ulem}
\usepackage[caption=false]{subfig}

\setlist[itemize]{align=parleft,left=0pt..1em}

\definecolor{Gray}{gray}{0.9}
\SetKwComment{Comment}{/* }{ */}

\newlength\myindent
\newcommand{\n}{GEM}
\newcommand{\nemb}{auxiliary}
\newcommand{\bisage}{BiSAGE}

\allowdisplaybreaks

\begin{document}

\title{Semi-supervised Learning with Network Embedding on Ambient RF Signals for Geofencing Services\vspace{-0.1in}}

\author[ ]{Weipeng Zhuo\textsuperscript{$\ast$}, Ka Ho Chiu\textsuperscript{$\ast$}, Jierun Chen\textsuperscript{$\ast$}, Jiajie Tan\textsuperscript{$\ast$} \\ Edmund Sumpena\textsuperscript{$\star$}, S.-H. Gary Chan\textsuperscript{$\ast, \S$}, Sangtae Ha\textsuperscript{$\dagger$},  Chul-Ho Lee\textsuperscript{$\ddagger, \S$}\vspace{-0.1in}}
\affil[$\ast$]{The Hong Kong University of Science and Technology, \textsuperscript{$\star$}Johns Hopkins University}
\affil[ ]{\textsuperscript{$\dagger$}University of Colorado Boulder, \textsuperscript{$\ddagger$}Texas State University \thanks{\hspace{-0.1in}\noindent This work was supported, in part, by Hong Kong General Research Fund (under grant number 16200120). The work of Chul-Ho Lee was supported, in part, by the NSF under Grant IIS-2209921. \vspace{1mm} \newline $^\S$Corresponding authors.}}

\maketitle

\begin{abstract}

In applications such as elderly care, dementia anti-wandering and pandemic control, it is important to ensure that people are within a predefined area for their safety and well-being. We propose \n{}, a practical, semi-supervised Geofencing system with network EMbedding, which is based only on ambient radio frequency (RF) signals. \n{} models measured RF signal records as a weighted bipartite graph. With access points on one side and signal records on the other, it is able to precisely capture the relationships between signal records. \n{} then learns node embeddings from the graph via a novel bipartite network embedding algorithm called \bisage{}, based on a \textbf{Bi}partite graph neural network with a novel bi-level \textbf{SA}mple and aggre\textbf{G}at\textbf{E} mechanism and non-uniform neighborhood sampling. Using the learned embeddings, \n{} finally builds a one-class classification model via an enhanced histogram-based algorithm for in-out detection, i.e., to detect whether the user is inside the area or not. This model also keeps on improving with newly collected signal records. We demonstrate through extensive experiments in diverse environments that \n{} shows state-of-the-art performance with up to 34\% improvement in $F$-score. \bisage{} in \n{} leads to a 54\% improvement in $F$-score, as compared to the one without \bisage{}.

\end{abstract}

\section{Introduction}
\label{sec:intro}
Many applications have been enabled by digital geofencing technology. For instance, in nursing homes for the elderly and patients~\cite{helmy2016alzimio} and medical observation for pandemic control~\cite{SignatureHome}, a user with an IoT device that can sense radio frequency (RF) signals stays within a designated area for a certain period of time. The user or the user's caregiver gets informed when the user gets out of the area. In constrained navigation of unmanned aerial vehicles (UAVs)~\cite{hermand2018constrained} and logistics management~\cite{oliveira2015intelligent}, a UAV or a freight is restricted to move within a specific region. Otherwise, an alert will be issued. In these cases, exact locations of the user within the area are irrelevant.

A traditional approach for geofencing is to localize a user with GPS or cell tower triangulation~\cite{greenwald2011economically,zang2010bayesian,nirjon2014coin,schloemann2015toward,rizk2018cellindeep}. However, it is \emph{ineffective} in indoor environments or highly complex metropolitan environments like New York due to non-line-of-sight radio propagation or signal fading. Another approach is to exploit a network-based localization~\cite{SignatureHome,margolies2017can}, which relies on the IP address of a user's network device. This approach may work for relatively large space but does not work well for indoor premises requiring house-level or room-level accuracy. Furthermore, indoor localization systems~\cite{hoang2019recurrent,li2019smartloc,wang2020spatial,zhou2021integrated,fan2021siabr,chen2022fidora} may be leveraged for geofencing. However, they usually require maps and collections of fingerprints, i.e., pairs of RF signals and corresponding location labels, in the localization area, which are often infeasible or labor-intensive to obtain. In other words, they would introduce privacy concerns and extra overhead in deployment, when they are used for geofencing.

To design a non-obtrusive yet highly accurate geofencing system, it is desirable to use ambient RF signals available in the region. Recent advances in machine learning can also be leveraged to that end. However, there are non-trivial technical challenges. While each RF signal record is often represented as a vector of pairs of sensed access points (APs) and received signal strength (RSS) values from them, the records need to be of \emph{identical} length to be used for building a learning model~\cite{HBOS,taxSupportVectorData2004,chalapathy2018anomaly, DeepSVDD18,goyal2020drocc,zaheer2020old,liznerski2021explainable}. In other words, the RF environment needs to be fairly static, which is far from reality.  The APs and recorded RSS values can vary spatially and temporally. Even on the same spot, they can differ due to environmental changes. APs could also be added or removed.

\begin{figure*}[t]
    \vspace{-0pt}
	\centering
	\includegraphics[width=0.8\textwidth]{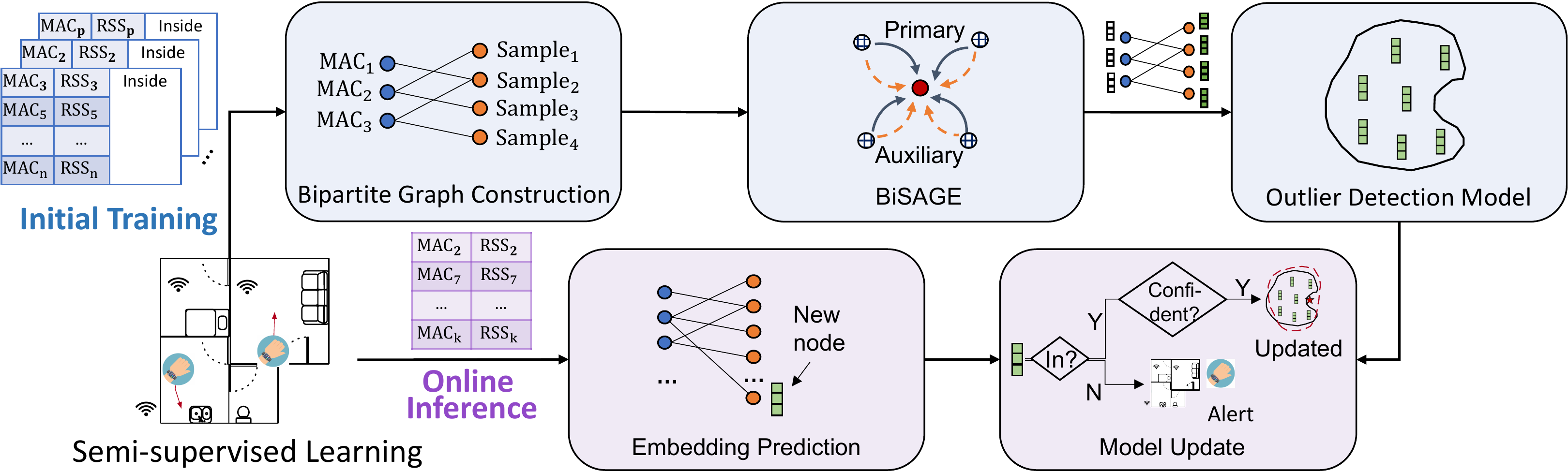}
	\vspace{-0.05in}
	\caption{A system overview of \n{}.}
	\vspace{-0.2in}
	\label{fig:sys_overview}
\end{figure*}

With the challenges in mind, we propose \textbf{\n{}} (\textbf{G}eofencing with network \textbf{EM}bedding), a semi-supervised learning system for automated IoT geofencing. As shown in Figure~\ref{fig:sys_overview}, in this system, a limited set of RF signal records, each having RSS values from ambient APs, are initially collected inside the geofencing area to build a one-class classification model for `in-out' detection, which is to detect whether the user is inside the area (normal) or outside (abnormal). A new RF signal record is then obtained on a regular basis and fed into the one-class classification model for the in-out detection. This model is built through the following three steps in \n{}.

From the initial signal records, \n{} first constructs \emph{a weighted bipartite graph}, reflecting the observation in each record while being \emph{dynamic} to include future incoming signal records. We represent an AP as a node of one type and a record \emph{itself} as a node of another type. Considering the inherent structure of each signal record having a list of sensed APs and their corresponding RSS values, we create edges between a node corresponding to each record and nodes corresponding to the APs detected in the record. We then assign weight values to the edges based on the RSS values. The connectivity structure in the bipartite graph not only reflects the information in each record but also captures the relevance between records. This graph representation quickly scales up to support a large number of incoming signal records over time. It is also flexible to reflect the arrivals (or departures) of new (or old) APs.

\n{} next extracts \emph{a low-dimensional vector representation} (node embedding) of each node from the weighted bipartite graph. It naturally avoids the issue with the signal records of variable size since the node embeddings are of equal length and they are used on behalf of the signal records. To this end, we propose \bisage{}, a novel  \textbf{Bi}partite network embedding algorithm with \textbf{SA}mple and aggre\textbf{G}at\textbf{E}. It adopts a graph neural network architecture inspired by GraphSAGE~\cite{graphsage}, while having its own algorithmic innovations tailor-made for \emph{weighted bipartite} graphs. Note that GraphSAGE was designed for \emph{homogeneous} graphs where all nodes are of the same type.  As shall be shown in Section~\ref{sec:exp}, we empirically demonstrate that \n{} (with \bisage{}) outperforms its version with GraphSAGE. We also discuss how \bisage{} is different from other bipartite network embedding algorithms in Section~\ref{sec:related}.

\begin{figure}
\centering
\includegraphics[width=0.7\linewidth]{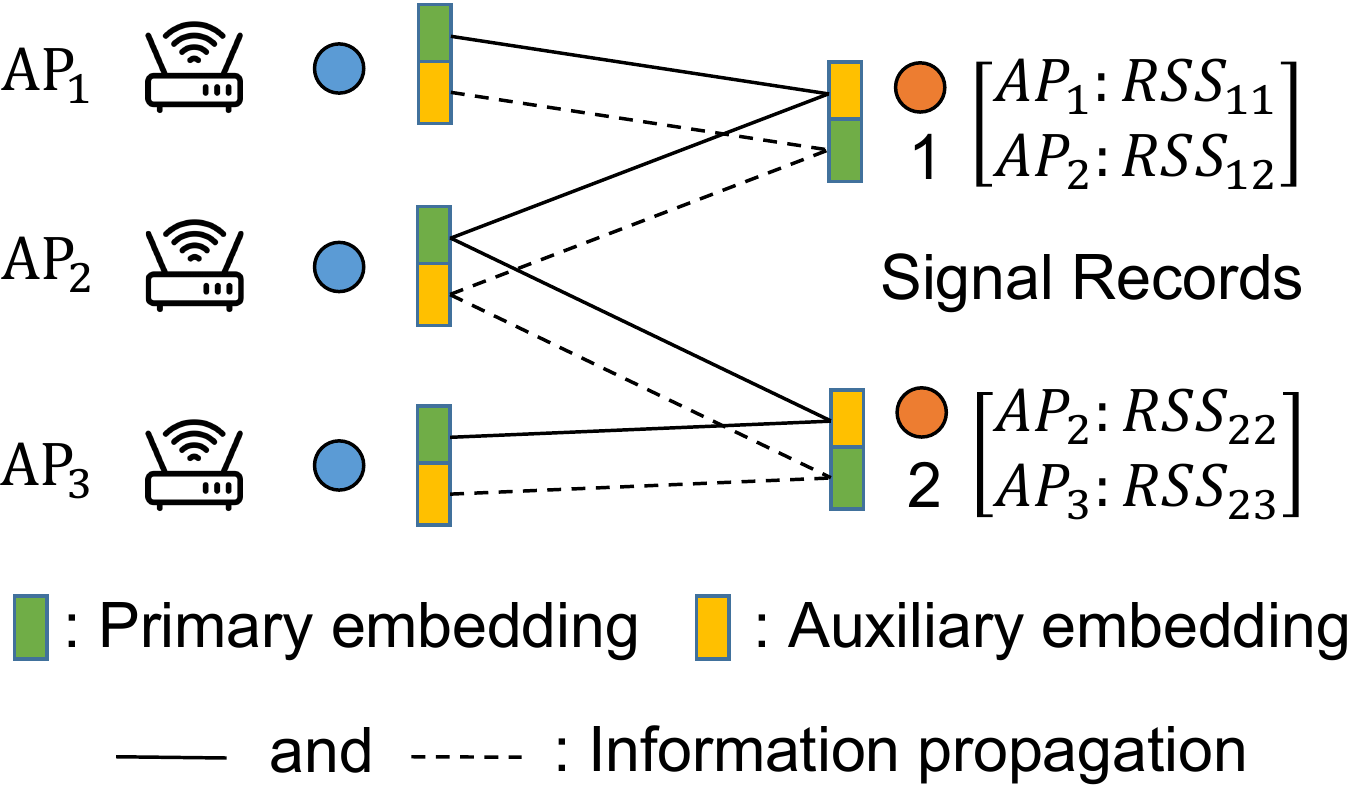}
\vspace{-0.05in}
\caption{In \bisage{}, the auxiliary embedding of a node serves as a carrier for information propagation of the primary embeddings of its neighbors (that are of the other node type) in the bipartite graph. The primary embeddings are the ones used for classification.}
\label{fig:intuition}
\vspace{-0.2in}
\end{figure}

Specifically, in \bisage{}, we first propose a novel \emph{bi-level aggregation} mechanism to differentiate embeddings for the nodes of different types and update the node embeddings based only on the ones from the nodes of the same type. To be precise, we introduce an \emph{auxiliary} embedding for each node in addition to its \emph{primary} node embedding. This auxiliary embedding is used as a `carrier' for information propagation in a way that carries an aggregation of primary node embeddings from the nodes of a different type than the current node and bypasses the current one. For example, when updating the primary embeddings of the `signal-record' nodes, the auxiliary embeddings of the `AP' nodes are passed to their neighboring signal-record nodes, but the AP nodes' primary embeddings are left untouched, and vice versa. As shown in Figure~\ref{fig:intuition}, the primary embeddings of signal-record nodes 1 and 2 can communicate with each other through the auxiliary embedding of AP node 2, without affecting the primary embedding of AP node 2. Second, since the graph is a weighted graph, we employ \emph{non-uniform} neighborhood sampling based on edge weights. Thus, a form of `attention' is naturally introduced in the neighborhood aggregation due to the edge weights. Third, we introduce a new loss function to obtain both the primary and auxiliary embeddings per node.

Once the embeddings of initial RF signal records are obtained via the weighted bipartite graph modeling and \bisage{}, they form a training set of `in-premises' (normal) data samples for an \emph{enhanced histogram-based} one-class classification algorithm for in-out detection. The rationale behind the histogram-based detection algorithm is to better capture the (potentially multimodal) distribution of `normal' data samples since the data samples, i.e., RF signal records, can exhibit quite different characteristics depending on where they are collected within the area. Furthermore, the training data samples are relatively small, so their representation of the geofencing area, especially its boundary, may be coarse. To cope with this, every incoming signal record (in its embedding) is used not only for in-out detection but also to augment the in-premises data for the histogram-based detection algorithm, if it is predicted to be highly confident as an `in-boundary' data sample. This (unsupervised) data augmentation continues over time and leads to better in-out detection performance.

In summary, we have the following contributions:

\vspace{1mm}

\begin{itemize}[itemsep=4pt,leftmargin=1em]

    \item \emph{\n{} effectively learns a vector representation of each RF record via our novel embedding algorithm \bisage{}}. Since each node in the weighted bipartite graph is transformed into a low-dimensional vector space via \bisage{}, the RF records of variable length are now mapped onto the same space, so their similarities and differences can be better identified. The node embeddings generated by \bisage{} also lead to better in-out detection performance, as compared to the ones by GraphSAGE.

    \item \emph{\n{} is self-evolving with new RF signal records.} Upon the arrival of a new RF signal record, our histogram-based detection algorithm is used not only for in-out detection but also to augment the `in-premises' data, if the new record is predicted to be a highly confident in-premise data sample. This self-enhancement makes the in-out detection performance keep on improving.

    \item \emph{\n{} significantly outperforms state-of-the-art algorithms.} We implement \n{} in Android phones and evaluate its performance under various housing and RF environments. Experiment results show that \n{} achieves state-of-the-art detection performance, thanks to its adaptivity to dynamic RF environments and its self-enhancement over time. With the bipartite graph modeling and \bisage{}, \n{} improves by 54\% in $F$-score over the case without them. \n{} outperforms state-of-the-art algorithms by up to 34\% in $F$-score. Furthermore, \n{} remains effective for a wide range of densities of ambient APs.

\end{itemize}

\section{Related Work}
\label{sec:related}
\noindent\textbf{Geofencing:} There are few recent studies that are closely relevant to this work. SignatureHome~\cite{SignatureHome} learns a geofencing area using network connectivity, such as association with a certain AP, and building a database of RF signal readings from surrounding APs. The RF signal records are converted into fixed-length vectors where missing entries are padded with arbitrarily small values to indicate nonavailability. In a similar vein, INOA~\cite{INOA18} leverages the RF signal readings from a set of APs for area classification. It changes each variable-size record into a set of records in which each record contains RSS values for each pair of sensed APs. INOA then builds a machine learning model based on the new records. However, our system \n{} does not require such a conversion of RF signal records but is able to use all the records regardless of how many RSS values are in each record, thanks to representing them in a bipartite graph and learning their (fixed-length) embeddings via \bisage{}. Furthermore, we empirically demonstrate that \n{} is \emph{superior} to SignatureHome and INOA.

Geofencing can also be considered as an indoor localization problem~\cite{boysen2014constructing,li2016vita,baba2016learning,hoang2019recurrent,li2019smartloc,wang2020spatial,lin2021locater,zhou2021integrated,fan2021siabr,chen2022fidora}. For example, Fidora~\cite{chen2022fidora} collects channel state information for each location in the geofencing area and trains a classifier to predict the location of a user. SIABR~\cite{fan2021siabr} builds a database of fingerprints, i.e., pairs of RF signals and corresponding location labels, to train bi-directional LSTM models for location inference. However, construction of such a database is labor-intensive and time-consuming. They also require indoor floorplans to obtain exact locations of users, which may raise privacy concerns. LOCATER~\cite{lin2021locater} first obtains a rough estimate of the location of the user, i.e., a predefined region covered by an AP, by recognizing the user's associated AP. It then detects which room the user is in by using the user's past behavior patterns and her potential companions (i.e., the appearance of their devices) recorded in the database. However, the user behavior analysis requires substantial logs from the past user activities, which are not available in our case. Moreover, since these indoor-localization methods merely provide an estimate of the user's location, the map of the geofencing area should be available for the location estimate to be usable for in-out detection. In contrast, our proposed system \n{} only requires RF signals collected inside the area, which is \emph{light-weight} and \emph{privacy-preserving}. It is also efficient in detecting outliers once an out-of-boundary event happens.

\vspace{2pt}
\noindent \textbf{Learning on heterogeneous graphs:} In \n{}, BiSAGE learns the embeddings of the nodes of a bipartite graph, which is a kind of heterogeneous graph. We here review recent network embedding and graph neural network algorithms for heterogeneous graphs. BiNE~\cite{gao2018bine} and E-LINE~\cite{GRAFICS} are network embedding algorithms for bipartite graphs, but are transductive learning algorithms to learn node embeddings from a bipartite yet \emph{static} graph. However, BiSAGE is an inductive learning algorithm, so it can be readily used for obtaining the embeddings of new nodes (e.g., new RF signal records) that are streamed and added into the bipartite graph. In addition, GRAPE~\cite{you2020handling} is a novel algorithm for representation learning on bipartite graphs. Its main focus is to deal with missing values in the features associated with nodes for representation learning, and it merely treats the bipartite graph as a homogeneous one. However, BiSAGE uses the underlying bipartite graph \emph{as is} to learn node embeddings from the graph without any \emph{prior} node features.

For learning on heterogeneous graphs (beyond the bipartite graphs), the concepts of `metapaths' have been used to guide the aggregation of information, e.g., features and embeddings, from the nodes of different types~\cite{dong2017metapath2vec,shi2018heterogeneous,fu2020magnn,chatzopoulos2020sphinx,fang2020effective,wang2020dynamic,yang2021interpretable,gu2022hybridgnn}. A metapath is an abstraction of a network path across the nodes of different types. For example, based on the metapath-based aggregation,
HybridGNN~\cite{gu2022hybridgnn} explores the importance of inter-relationship between different nodes through a randomized exploration to learn better node embeddings. However, when it comes to bipartite graphs, it is deemed \emph{unnecessary} since there are only two types of nodes. In addition, attention mechanisms have been used to implicitly specify weights to edges into the nodes of different types~\cite{wang2019heterogeneous,yang2020multisage,huang2022estimating}. For example, HIVEN~\cite{huang2022estimating} is developed for a heterogeneous graph where each node may have multiple types of relationship with others. It learns type-wise embeddings of each node, which are then combined to obtain the final node embedding per node using an attention mechanism where the attention weights are learned through training. In contrast, by definition, our underlying graph is a \emph{weighted} graph, which does not necessarily require learning the attention weights. Furthermore, BiSAGE enables \n{} to achieve (almost) ideal performance without such an additional mechanism.

\vspace{2pt}
\noindent{\textbf{Outlier detection:}} The in-out detection from the geofencing area can be cast as an outlier detection~\cite{kdd17Anomaly,DeepSVDD18,corain2021dbscout,kieu2022robust} problem where any RF signal records obtained \emph{outside} the area are to be detected as outliers. DBSCOUT~\cite{corain2021dbscout}, which is built on DBSCAN, treats a point as an outlier if it is detected outside the hypersphere (dense region) with a predefined radius. It is, however, tested only on two- and three-dimensional data. RDAE~\cite{kieu2022robust} leverages two layers of autoencoders and decomposes input data into normal and outlier data samples for training. In contrast, \n{}, especially our enhanced histogram-based detection algorithm, does not require any \emph{labeled} outliers as input data samples, but is able to detect outliers using the training dataset that consists of only normal data samples. Furthermore, its performance is extensively evaluated in comparison with several popular outlier detection algorithms, such as feature bagging, isolation forest, and local outlier factor.

\section{\n{}: Initial Training}
\label{sec:offline}
In this section, we explain the details of the three integral components of \n{} to build a one-class classification model.

\subsection{Weighted Bipartite Graph Model}
\label{subsec:bipartite}
It has been a common practice to represent the RF signal data in a matrix form (or vectors of equal length), when they are leveraged for various applications such as indoor localization~\cite{abbas2019wideep,li2021spatial} and floor identification~\cite{kim2018scalable}. Each signal record (sample) is a vector of RSS values from surrounding APs that are indicated by their medium access control (MAC) addresses and becomes a column of the matrix.\footnote{Note that each AP can have one or more MAC addresses associated with its transceivers, and the MAC addresses are the ones actually recorded in each RF signal record. We use MAC addresses instead of APs for clarity.} In other words, each RSS value associated with a different MAC address is for a row of the matrix, so each row index corresponds to a different MAC address. In the matrix representation, however, there is a `missing-value' problem in that some entries in the matrix lack RSS values, as the samples are of variable length (i.e., the number of detected MAC addresses can be different per sample). While the missing entries are generally filled with arbitrarily small values, this ad-hoc data imputation potentially leads to incorrect representation.

On the other hand, as recently used in~\cite{GRAFICS} for a floor classification application, the variable-length RF signal records can be modeled as a weighted bipartite graph, where the measured signal information is preserved \emph{without} any ad-hoc data imputation. Specifically, for each RF signal record, this record itself becomes a node of one type and the sensed MAC addresses in the record become nodes of the other type, with undirected edges connecting the `record' node and the `MAC' nodes. Each edge is assigned a weight that is determined as a function of the RSS value from its corresponding MAC address in the record.

\begin{figure}[t]
\vspace{1mm}
\centering
	\includegraphics[width=0.45\textwidth]{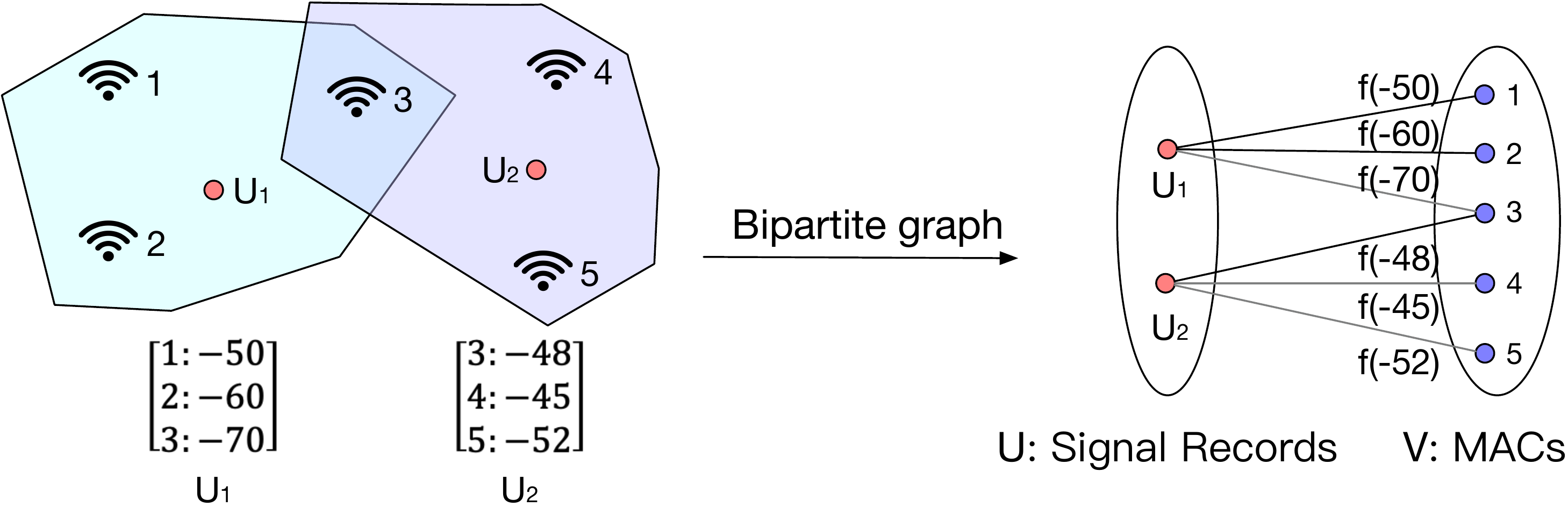}
 	\vspace{-0.1in}
	\caption{An illustrating example for two sensing events with five ambient APs.}
	\label{fig:bigraph}
 	\vspace{-0.2in}
\end{figure}

Observe that each RF signal record has a set of pairs of sensed MAC addresses and their corresponding RSS values. For each record (sample) $u$, we denote by $\text{RSS}_{uv}$ the RSS value from a MAC address, say $\text{MAC}_v$, which appears in the record. Let $U$ be a set of nodes for the signal records and $V$ be a set of nodes for the sensed MAC addresses. We then define a weighted bipartite graph $\mathcal{G} \!=\! (U, V, \mathcal{E}, \bm{w})$, where $\mathcal{E}$ is a set of edges with edge $e_{uv} \!\in\! \mathcal{E}$ denoting the edge between $u \!\in\! U$ and $v \!\in\! V$, and $\bm{w}$ is a set of edge weights with weight $w_{uv} \!\in\! \bm{w}$ of edge $e_{uv}$, which is defined as
\begin{equation}
w_{uv} := f(\text{RSS}_{uv}),
\label{eqn:edge_weight}
\end{equation}
where $f$ is a function of $\text{RSS}_{uv}$ such that $f(\text{RSS}_{uv}) \!>\! 0$ for all $\text{RSS}_{uv}$. Note that edge $e_{uv}$ indicates the presence of $\text{MAC}_v$ in record $u$. In other words, two nodes $u$ and $v$ are connected if there exists a measured value of $\text{RSS}_{uv}$ between them, with edge weight $w_{uv}$. In the example illustrated in Figure~\ref{fig:bigraph}, RF signal record $u_1$ is only connected to MACs 1--3, while $u_2$ is connected to MACs 3--5.

In \n{}, we also use the following weight function:
\begin{equation}
f(\text{RSS}_{uv}) := \text{RSS}_{uv} + c,
\label{eq:offset}
\end{equation}
where $c$ is a constant such that $c \!>\! \max\{|\text{RSS}_{uv}|, \forall u, v\}$~\cite{GRAFICS}. It is also worth noting that this bipartite graph representation can be easily extended to include the arrivals of new RF signal records and the changes in ambient APs (or their MAC addresses).

\begin{figure*}[t]
    \centering
    \includegraphics[width=0.8\textwidth]{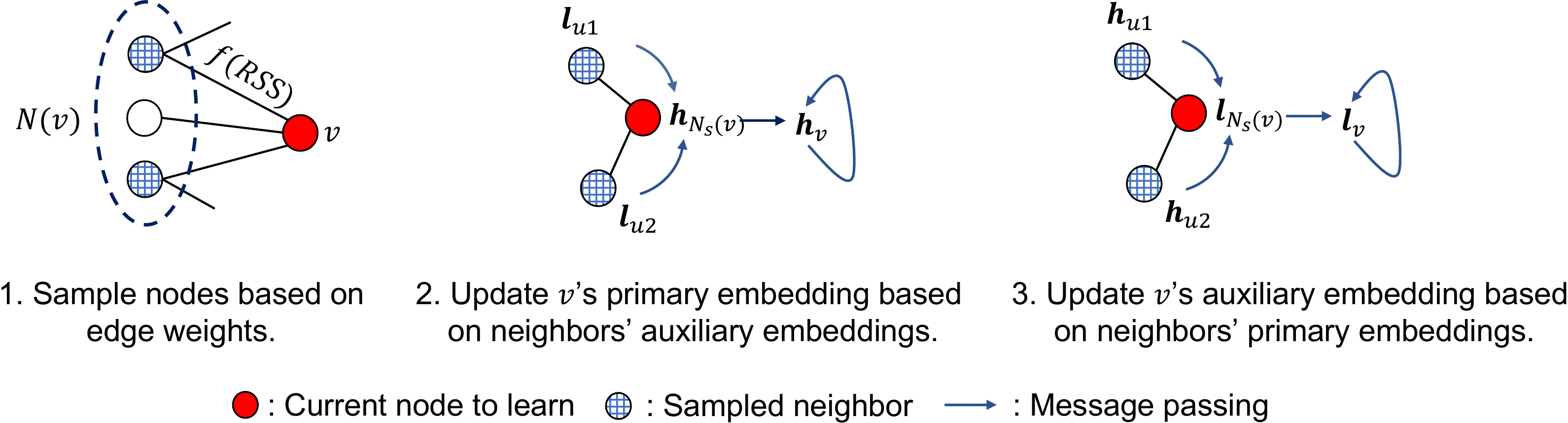}
    \vspace{-0.05in}
    \caption{To learn a node's embedding, \bisage{} first samples its neighbors and then aggregates the information from them via their auxiliary embeddings.}
    \label{fig:bi-level}
    \vspace{-0.15in}
\end{figure*}

\subsection{\bisage{}}
\label{subsec:bisage}

From the weighted bipartite graph $\mathcal{G}$, we next learn node embeddings of equal length to be used to build a one-class classification model for in-out detection. To this end, we propose \bisage{}, which is a non-trivial extension to GraphSAGE~\cite{graphsage} with a novel bi-level aggregation and a non-uniform neighborhood sampling.

Given a target node whose embedding is to be learned, there are two steps in the aggregation of embeddings from its (possibly multi-hop) neighbors. We sample nodes from the neighborhood of the target node according to a given probability distribution. Then, we aggregate the information (embeddings) from sampled neighboring nodes towards the target node (to update its embedding).

In the bipartite graph, edges are assigned weights that are a function of sensed RSS values. Thus, to determine which neighbors to sample, intuitively, the higher the sensed RSS value between the node and its neighbor, the more likely the neighbor should be chosen for the information aggregation. Thus, we design a non-uniform neighbor sampling based on edge weights.
Without loss of generality, suppose that $u \!\in\! U$ is the target node to learn its embedding. Let $N(u)$ be the set of neighbors of $u$ in the bipartite graph. The probability that $v \!\in\! N(u)$ is chosen for the aggregator function is defined as
\begin{equation*}
\Pr(v) = \frac{w_{uv}}{\sum_{v'\in N(u)} w_{uv'}}.
\end{equation*}

After sampling the neighbors of the target node, we need to aggregate the information from the sampled neighbors to the target node. If the underlying graph is a homogeneous graph, the information aggregation can be done for each node in the same way~\cite{graphsage}. However, in the bipartite graph (a heterogeneous graph), nodes of different types should be processed differently. Thus, we propose a novel bi-level aggregation mechanism for bipartite graphs.

In the bi-level aggregation, we introduce a new `auxiliary' embedding for each node. We consider the original embedding of each node as its `primary' embedding. In other words, each node has two embeddings, namely primary embedding and \nemb{} embedding. The \nemb{} embedding at each node now serves as a `carrier' for the information propagation when the target node is of different type. For example, if the target node is a `signal-record' node, its neighbors -- `sensed MAC' nodes -- use their \nemb{} embeddings to propagate the information.

As shown in Figure~\ref{fig:bi-level}, for the target node, we sample its neighbors based on their edge weights for the information aggregation. The target node's primary embedding is then updated based on its sampled neighbors' \nemb{} embeddings that carry the information (embeddings) from the nodes of the same type. On the other hand, the target node's \nemb{} embedding is updated based on its sampled neighbors' primary embeddings to carry the information. Note that the \emph{auxiliary} embedding at each node constrains the influence of the information within the nodes of the same type, while the information can still flow over the graph.

\begin{figure*}[t]
    \centering
    \includegraphics[width=0.9\textwidth]{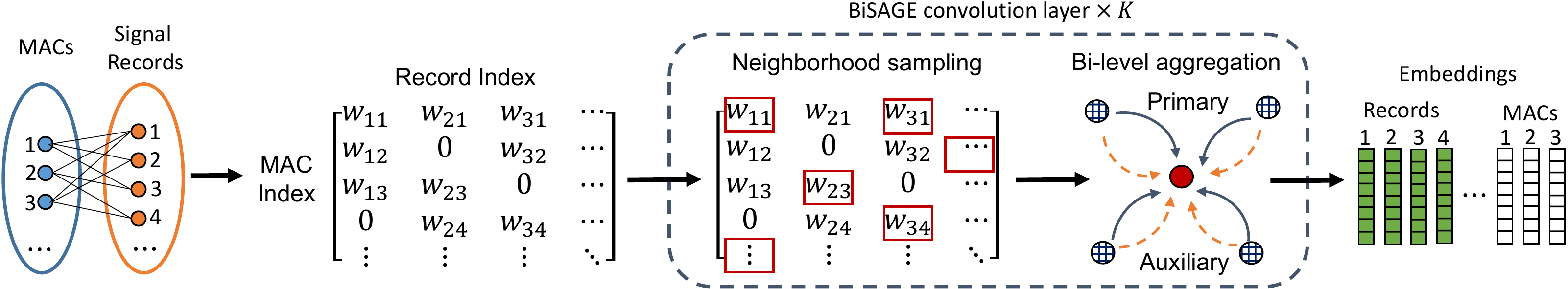}
    \vspace{-0.05in}
    \caption{An overall procedure from constructing a weighted bipartite graph to obtaining node embeddings via BiSAGE.}
    \label{fig:gnn_bi}
    \vspace{-0.1in}
\end{figure*}

For node $i \!\in\! U \!\cup\! V$, let $\bm{h}_i$ and $\bm{l}_i$ be its $d$-dimensional primary and \nemb{} embeddings, respectively. We denote by $N_s(i)$ the \emph{sampled} neighborhood of $i$. Let $k$ be the current aggregation step and $K$ be the total number of steps. In the $k$-th round of the aggregation, we need to update both primary and \nemb{} embeddings for each node. To update the primary embedding of $i$, we need to aggregate the information through its neighbors' \nemb{} embeddings. Thus, we have
\begin{align}
\bm{h}_{N_s(i)}^k &\longleftarrow \text{AGGREGATE}({\bm{l}_j^{k-1}, \forall j \in N_s(i)}), \label{eqn:agg_primary0}\\
\bm{h}_i^k &\longleftarrow \sigma(\bm{W}^k_h\cdot \text{CONCAT}(\bm{h}_i^{k-1}, \bm{h}_{N_s(i)}^k)), \label{eqn:agg_primary}
\end{align}
where $\text{AGGREGATE}(\cdot)$ is an aggregator function, e.g., MEAN($\cdot$) or MAX($\cdot$), $\bm{h}_{N_s(i)}$ is a $d$-dimensional temporary vector to store the aggregation result from the sampled neighborhood of $i$, $\text{CONCAT}(\cdot)$ is a concatenation function, $\bm{W}^k_h$ is a learnable weight matrix for primary embedding, and $\sigma$ is a nonlinear function. Here we introduce the superscript $k$ to indicate the $k$-th round of the aggregation. Similarly, to update the \nemb{} embedding of $i$, we need to aggregate the information from its neighbors' primary embeddings. Thus, we have
\begin{align}
\bm{l}_{N_s(i)}^k &\longleftarrow \text{AGGREGATE}({\bm{h}_j^{k-1}, \forall j \in N_s(i)}), \label{eqn:agg_auxiliary0}\\
\bm{l}_i^k &\longleftarrow \sigma(\bm{W}^k_l\cdot \text{CONCAT}(\bm{l}_i^{k-1}, \bm{l}_{N_s(i)}^k)),
\label{eqn:agg_auxiliary}
\end{align}
where $\bm{l}_{N_s(i)}$ is another temporary vector to store the aggregation result from the sampled neighborhood of $i$, and $\bm{W}^k_l$ is another learnable weight matrix for the auxiliary embedding.

After completing the $k$-th round of the aggregation, $h_i^k$ and $\bm{l}_i^k$ are then normalized by their $\ell_2$ norms, respectively, as follows:
\begin{equation}
    \bm{h}_i^k = \frac{\bm{h}_i^k}{||\bm{h}_i^k||_2}, ~\text{and}~~  \bm{l}_i^k = \frac{\bm{l}_i^k}{||\bm{l}_i^k||_2},    \label{eqn:agg_normalize}
\end{equation}
which are the embeddings to be used for $(k\!+\!1)$-th iteration. After $K$ iterations, the final primary and \nemb{} embeddings of node $i$ are $\bm{h}_i^K$ and $\bm{l}_i^K$, respectively. Initially, $\bm{h}_i^0$ and $\bm{l}_i^0$ are chosen randomly.

For our choice of the aggregator function, we observe that the higher the edge weight (based on the RSS value), the closer the two nodes having the edge should be. In other words, the information propagating through edges with higher weights should be considered more important during the aggregation. Thus, we use the following weighted aggregator function for $\bm{h}_{N_s(i)}^k$ in Equation~\eqref{eqn:agg_primary0}:
\begin{equation}
\bm{h}_{N_s(i)}^k := \sum_{j\in N_s(i)}\left(\frac{w_{ji}}{\sum_{j'\in N_s(i)} w_{j'i}}\bm{l}_j^{k-1}\right).
\label{eqn:weighted_agg}
\end{equation}
Similarly for $\bm{l}_{N_s(i)}^k$ in Equation~\eqref{eqn:agg_auxiliary0}.

Next, we explain the details of the training process. A standard approach to training a model for network embedding and representation learning on a graph is to leverage random walks. They are used to learn embeddings such that the nodes that appear in the sequence of nodes visited by the same walk should be close to each other in the embedding space~\cite{DeepWalk,graphsage}. Let $\{X_k\}^n_{k=0}$ be a finite sequence of nodes visited by a random walk when it starts from node $x_0 \!\in\! U \cup V$, i.e., $X_0 \!=\! x_0$. Given the current node of the random walk, say $X_k \!=\! x_k$, the next node $x_{k+1}$ is chosen from the neighbors of $x_k$ according to a certain distribution. As in GraphSAGE~\cite{graphsage}, a popular example is to use a uniform distribution for choosing the next node, i.e., the next node is chosen from the neighbors of the current node \emph{uniformly at random}. However, our graph is a \emph{weighted} (bipartite) graph, where each edge weight is proportional to its corresponding measured RSS value, implying that a higher RSS value generally indicates a closer distance between the IoT device and the AP. Thus, in BiSAGE, we use a weighted random walk, whose transition probabilities are given by
\begin{equation*}
    \Pr\left(X_{k+1} = x_{k+1} |  X_k = x_k \right) = \frac{w_{x_k x_{k+1}}}{\sum_{x' \in N(x_k)} w_{x_k x'}}.
\end{equation*}

In addition, we need to specify a loss function to minimize in learning the two sets of weight matrices $\{\bm{W}^k_h\}^K_{k=1}$ and $\{\bm{W}^k_l\}^K_{k=1}$ for primary and auxiliary embeddings, respectively. The rationale behind our loss function is that the primary embedding of one node should be close to the \nemb{} embedding of its neighbor visited by the random walk, as they encode the information of nodes of the same type, and vice versa. Let $x$ and $y$ be two neighboring nodes that appear as the nodes consecutively visited by the random walk. Let  $\bm{h}_{x}$, $\bm{l}_{x}$, $\bm{h}_{y}$, and $\bm{l}_{y}$ be their corresponding primary and \nemb{} embeddings, respectively. Specifically, we use the following loss function to minimize in learning the weight matrices $\{\bm{W}^k_h\}$ and $\{\bm{W}^k_l\}$ to be used to obtain the primary and \nemb{} embeddings of each node:
\begin{align}
    J_{\mathcal{G}} &:= - \log \left[ \sigma (\bm{h}_{x} \cdot \bm{l}_{y}) \sigma(\bm{l}_{x} \cdot \bm{h}_{y}) \right] \nonumber \\
    &- K_N \mathbb{E}_{z \sim \Pr(z)} \left[\log \left[ \sigma(- \bm{h}_{x} \cdot \bm{l}_{z}) \sigma (-\bm{l}_{x} \cdot \bm{h}_{z}) \right] \right],\label{eqn:loss}
\end{align}
where $\bm{a} \cdot \bm{b}$ is the inner product between $\bm{a}$ and $\bm{b}$, $\sigma(x) \!:=\! 1/(1+\exp(-x))$, $K_N$ is the number of `negative' samples, and the expectation $\mathbb{E}$ is with respect to $z$ drawn according to $\Pr(z)$, $z \in U \cup V$.

Our loss function in Equation~\eqref{eqn:loss} is based on the negative sampling technique~\cite{Mikolov2013,graphsage}, but defined in a way that is suitable for our bi-level aggregation on a weighted bipartite graph. The first term of the loss function encourages nodes of the same type that co-occur in the same random walk to stay close to each other in the embedding space. Recall that the \nemb{} embedding of node $y$ encodes the information of its neighbors, which have the same type as node $x$. On the other hand, the second term uses $K_N$ (negative) sampled nodes drawn from the graph and forces them to separate apart from node $x$, since the sampled nodes are more likely far from the current node $x$. We use $K_N = 4$ and $\Pr(z) \propto \text{deg}_z^{3/4}$, where $\text{deg}_z$ is the degree of node $z$~\cite{Mikolov2013,graphsage} .

\begin{figure}[t]
	\centering
	\includegraphics[width=0.4\textwidth]{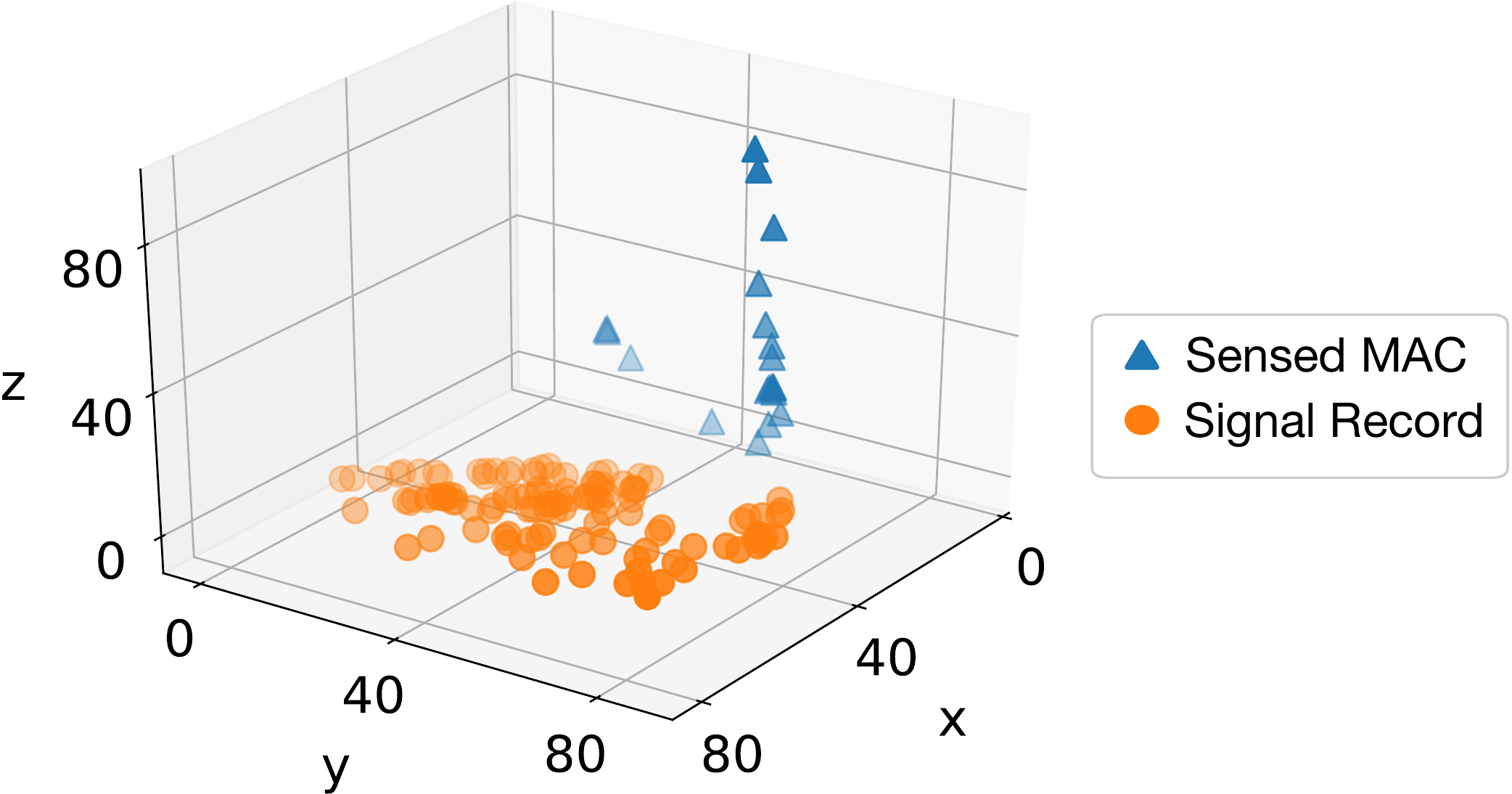}
    \vspace{-0.1in}
	\caption{A visualization of our learned embeddings.}
	\label{fig:ap_obs}
    \vspace{-0.2in}
\end{figure}

In Figure~\ref{fig:gnn_bi}, we summarize the overall process from constructing a weighted bipartite graph to obtaining node embeddings of the graph via \bisage{}. To show the effectiveness of our learned node embeddings, we use a visualization tool called t-SNE~\cite{van2008visualizing} to visualize the embeddings. We collect RF signal records inside a room and process them with \bisage{}. In general, signal records collected nearby would experience similar fading conditions, including multipath fading effects, though not exactly the same. Our \bisage{} is able to capture such a similarity efficiently. As shown in Figure~\ref{fig:ap_obs}, nodes of the same type generally stay together while being separated from nodes of different types. In addition, as shall be validated in Section~\ref{sec:exp}, the learned embeddings preserve the relevance among the nodes of the same type, i.e., their distance in the embedding space preserves their physical distance, which allows us to detect outliers based on their proximity information.

The bi-level aggregation mechanism is summarized in Algorithm~\ref{algo:bisage}. Its time complexity can also be analyzed as follows. Let $N_s$ be the number of sampled neighbors for each node, i.e., $N_s \!=\! |N_s(i)|$ for all $i$. Note that it is a hyperparameter, whose value is normally chosen between 10 and 25. Lines 4 and 5 take $O(d N_s)$ operations each, where $d$ is the embedding dimension. Here we use the weighted average in Equation~\eqref{eqn:weighted_agg} as an aggregate function. Each of Lines 6 and 7 involves a matrix multiplication with $\bm{W}^k\in \mathbb{R}^{d\times 2d}$, which has the complexity of $O(d^2)$. Lines 8 and 9 perform normalization of a $d$-dimensional embedding each, which has the complexity of $O(d)$. Therefore, the time complexity of Algorithm~\ref{algo:bisage} is $O\left(K(|U|\!+\!|V|)(dN_s\!+\!d^2)\right)$. It is worth noting that the matrix multiplication is often done in parallel on GPUs, and we also use the minibatch training strategy as proposed in GraphSAGE~\cite{graphsage}. Thus, the complexity would be much lower in practice.

\begin{algorithm}[hbt!]

\caption{\bisage{}: Bi-level Aggregation.}\label{algo:bisage}

\textbf{Input:} Bipartite graph $\mathcal{G}$; Total number of aggregation layers $K$. 

\textbf{Output:} Primary and auxiliary embeddings $\bm{h}_i$ and $ \bm{l}_i$ for all $i \in U \cup V$. 

\nl  Initialize $\bm{h}_i$ and $\bm{l}_i$ for all $i \in U \cup V$.

\nl \For{$k= 1, 2, \ldots, K$}{
\nl     \For{$i \in U \cup V$} {
        \Comment{Neighborhood aggregation}
\nl        $\bm{h}_{N_s(i)}^k := \text{AGGREGATE}({\bm{l}_j^{k-1}, \forall j \in N_s(i)}) $
        
\nl        $\bm{l}_{N_s(i)}^k := \text{AGGREGATE}({\bm{h}_j^{k-1}, \forall j \in N_s(i)}) $

        \Comment{Embedding update}
\nl        $\bm{h}_i^k := \sigma(\bm{W}^k_h\cdot \text{CONCAT}(\bm{h}_i^{k-1}, \bm{h}_{N_s(i)}^k))$
        
\nl        $\bm{l}_i^k := \sigma(\bm{W}^k_l\cdot \text{CONCAT}(\bm{l}_i^{k-1}, \bm{l}_{N_s(i)}^k))$

        \Comment{Normalization}
\nl        $\bm{h}_i^k := \frac{\bm{h}_i^k}{||\bm{h}_i^k||_2}$  
        
\nl        $\bm{l}_i^k := \frac{\bm{l}_i^k}{||\bm{l}_i^k||_2}$
     }
}
\nl \Return{$\bm{h}_i, \bm{l}_i,~i \in U \cup V$.}

\end{algorithm}

\subsection{In-Out Detection}
\label{subsec:boundary_training}

We next explain our enhanced histogram-based outlier detection model that is built on the primary embeddings of RF signal records for in-out detection, to detect whether the user is inside the area (normal) or outside (outlier). Specifically, we adopt and \emph{enhance} the histogram-based algorithm~\cite{HBOS} due to its simplicity (fast training) and effectiveness in capturing a hidden distribution in the feature space.\footnote{Recall that RF signal records can be quite different depending on where they are collected within the area, thereby making the distribution potentially multimodal. A histogram-based approach is suited for capturing the multimodality in the distribution.} It leverages the idea that the feature vectors from normal data exhibit similar patterns in the feature space, as long as they well represent the data. For each feature (each dimension of the feature vector), it builds a histogram of frequencies obtained from the feature vectors of training (normal) data.

We below explain the details of the histogram-based algorithm when used with the primary embeddings of signal records (initial normal data), and then present our enhancement. Suppose that there are $n$ $d$-dimensional primary embeddings $\bm{h}_1, \bm{h}_2, \ldots, \bm{h}_n$. Let $h_{i,j}$ be the $j$-th element (or dimension) of the $i$-th embedding. We first obtain the maximum value, i.e., $\Delta^u_j \!:=\! \max_i\{h_{i,j}\}$, and the minimum value, i.e., $\Delta^l_j \!:=\! \min_i\{h_{i,j}\}$, from the $j$-th element values of the $n$ embeddings. We then create $m$ bins, with each bin of width $(\Delta^u_j-\Delta^l_j)/m$. Since the $j$-th element value of each embedding $\bm{h}_{i}$ falls into one of the bins, we can construct a histogram based on the frequency count of each bin. Note that there is a separate histogram for each dimension $j$.

Given the $d$ histograms, when a new (primary) embedding is available, i.e., a new RF signal record is available and its corresponding primary embedding is obtained (via BiSAGE), we can compute its `outlier' score and determine whether it is an outlier by comparing the score with a threshold value. The score function and the threshold value are given as follows. For any given embedding, say $\bm{h} \!:=\! [h_1, h_2, \ldots, h_d]$, its `outlier' score is calculated by
\begin{equation}
\setlength{\abovedisplayskip}{3pt}
\setlength{\belowdisplayskip}{3pt}
H(\bm{h}) = \sum_{j=1}^{d} \log \left(\frac{1}{\text{hist}_j(h_j)} \right),
\label{eqn:histogram}
\end{equation}
where $\text{hist}_j(h_j)$ is the frequency count of the bin where $h_j$ belongs for each dimension $j$. The \emph{higher} the score, the more likely it is an outlier. As for the threshold value, we compute the outlier scores of the $n$ primary embeddings of the initial normal data using Equation~\eqref{eqn:histogram}. We then use the min-max normalization to normalize the scores into the range $[0,1]$, and sort them in decreasing order. To be precise, let $\bar{H}(\bm{h}_1), \bar{H}(\bm{h}_2), \ldots, \bar{H}(\bm{h}_n)$ denote the normalized scores, and let $\bar{H}(\bm{h}_{[1]}), \bar{H}(\bm{h}_{[2]}), \ldots, \bar{H}(\bm{h}_{[n]})$ denote the \emph{sorted}, normalized scores, where the subscript $[i]$ indicates the index of an embedding that leads to the $i$-th highest score. With $\gamma$ defined as a `contamination' factor and $i^* \!:=\! n\times \gamma$, the threshold value, denoted by $\tau$, for outlier detection is originally set to $\tau := \bar{H}(\bm{h}_{[i^*]})$~\cite{HBOS}.

While the contamination factor $\gamma$ can be used to control the level of the sensitivity in detecting outliers, the threshold value $\tau$ is highly dependent on the (initial) data size $n$ and the choice of $\gamma$. In our geofencing system, we aim to augment the normal data with newly measured RF records. Since the data size keeps on increasing, however, it can lead to changes in the threshold value $\tau$ and negatively affect the performance of outlier detection. For example, if the scores of recent records are somehow relatively much higher than the old ones, then they may boost up the threshold value $\tau$, which would lead to misclassifications  of outlier (outside) records as normal (in-premises) ones, and vice versa. Thus, there is a need for an enhanced algorithm to make this outlier detection adaptive in incorporating new records.

We first reduce the dependence of choosing the threshold value $\tau$ on the data size. We also observe that if the outlier scores for normal and abnormal records can be smoothed and separated further apart, it would be easier to set the threshold value. Thus, we adopt to use the softmax function with a scaling factor $T$. For any given embedding $\bm{h}$, its outlier score is now obtained by
\begin{equation}
\setlength{\abovedisplayskip}{4pt}
\setlength{\belowdisplayskip}{4pt}
S_T(\bm{h}) := \frac{\exp(\bar{H}(\bm{h})/T)}{\exp(\bar{H}(\bm{h})/T) + \exp((1-\bar{H}(\bm{h}))/T)}.
\label{eqn:enhanced_score}
\end{equation}
Note that $S_T$ is in the form of a Boltzmann distribution (or Gibbs distribution) in statistical physics, where $T$ is the thermodynamic temperature.

The intuition behind the transformation with the softmax function is that the outlier scores of \emph{normal} samples (i.e., the embeddings of RF signal records obtained \emph{inside} the geofencing area) get quite close to each other. Similarly for \emph{abnormal} samples, which are the embeddings of the ones measured \emph{outside} the area, including those collected just outside the boundary (see Section~\ref{sec:discuss} for further discussion on such abnormal samples). In addition, the scaling factor $T$ further separates the former apart from the latter. Therefore, we use the new score function $S_T$ in Equation~\eqref{eqn:enhanced_score} to compute the outlier score of a new sample $\bm{h}$, i.e., the (primary) embedding of a new RF signal record, and determine it is an outlier if
\begin{equation}
\setlength{\abovedisplayskip}{4pt}
\setlength{\belowdisplayskip}{4pt}
S_T(\bm{h}) > \tau_u,
\label{eqn:thresh}
\end{equation}
where $\tau_u$ is our new threshold value. The scaling parameter $T$ and the new threshold value $\tau_u$ are considered as hyperparameters to be optimized in the learning process. As shall be shown in Section~\ref{sec:exp}, we empirically demonstrate that the rescaling method  with the softmax function in Equation~\eqref{eqn:enhanced_score} for the histogram-based outlier detection indeed improves the in-out detection performance. 

\section{\n{}: Online Inference and Update}
\label{sec:online}
Once our outlier detection model is built on the primary embeddings of normal RF signal records (initial training data), \n{} performs in-out detection (or outlier detection) whenever a new RF signal record is available. The IoT device of the user periodically records RSS values from ambient APs, and they are fed into \n{} for the in-out detection. In what follows, we first explain how to obtain the primary embeddings of the new RF signal records in an online manner, and then present how to use the new records to perform in-out prediction and further update our outlier detection model over time.

\subsection{Embedding Prediction}
\label{subsec:graph_update}

The primary and \nemb{} embeddings of every node in the weighted bipartite graph are learned in the training process. When a new RF signal record becomes available, it is added into the graph as a new `signal-record' node, say $r$, and its edges are created with the `MAC' nodes that appear in the record. Some MAC nodes may also be newly added, if they are new ones to the graph. Then, edge weights are determined based on recorded RSS values according to Equations~\eqref{eqn:edge_weight} and~\eqref{eq:offset}.\footnote{It is possible that the new signal record only contains the MAC addresses that have \emph{never} been seen before. In this case, it may be the one collected far from the geofencing area, and thus we treat it as an outlier, raising an alert.} Once node $r$ is added into the graph, its primary embedding $\bm{h}_r$ and \nemb{} embedding $\bm{l}_r$ are obtained according to Equations~\eqref{eqn:agg_primary0}--\eqref{eqn:agg_normalize}, along with the learned weight matrices $\{\bm{W}^k_h\}$ and $\{\bm{W}^k_l\}$. In other words, node $r$ aggregates the information (embeddings) from its sampled neighbors, and this aggregation process is repeated $K$ times. That is, node $r$ is able to attain the knowledge from all nodes within the $K$-hop neighborhood, and thus its embeddings are determined by their embeddings.

\subsection{In-out Prediction}
\label{subsec:inout_prediction}
Given the learned primary embedding $\bm{h}_r$ of the new signal record $r$, we need to decide whether it is an outlier or not. To this end, we first calculate its raw outlier score $H(\bm{h}_r)$ as in Equation~\eqref{eqn:histogram} and normalize it through the min-max normalization to obtain the normalized score $\bar{H}(\bm{h}_r)$ (see Section~\ref{subsec:boundary_training} for more details). We then obtain its final outlier score $S_T(\bm{h}_r)$ as in Equation~\eqref{eqn:enhanced_score} and decide whether it is an outlier or not according to Equation~\eqref{eqn:thresh}.

\begin{algorithm}[hbt!]

\caption{\n{}: Online Inference.}\label{algo:inference}

\textbf{Input:} New signal record $\bm{r}$; Bipartite graph $\mathcal{G}$. 

\textbf{Output:} \textit{IN} or \textit{OUT} for $\bm{r}$. 

\nl Connect $\bm{r}$ into $\mathcal{G}$. 

\nl Obtain $\bm{r}$'s primary embedding $\bm{h}_r$ according to Equations~\eqref{eqn:agg_primary0}--\eqref{eqn:agg_normalize}.

\nl Calculate outlier score $S_T(\bm{h}_r)$ as in Equation~\eqref{eqn:enhanced_score}.

\nl \eIf{$S_T(\bm{h}_r) > \tau_u$} {
\nl     \Return {OUT}
}{
\nl     \If{$S_T(\bm{h}_r) < \tau_l$}{
\nl        Update $d$ histograms.   
    }    
\nl    \Return {IN}
}

\end{algorithm}

\subsection{Online Model Update}
\label{subsec:b_update}

As illustrated in Figure~\ref{fig:sys_overview}, if a new signal record is predicted to be an outlier by our outlier detection model, it triggers an alert. On the other hand, if it is predicted to be a normal one -- the record measured inside the geofencing area, we further check how confident the prediction is. Then, if it can be considered as a highly confident normal (in-premises) sample, we update our outlier detection model.

To this end, we introduce another yet more strict threshold value $\tau_l$ ($< \tau_u$) to filter out the `normal' predictions with \emph{low} confidence. Specifically, when the score of a new sample $\bm{h}_r$ -- the primary embedding of a new record -- is less than the threshold value $\tau_u$ (see Equation~\eqref{eqn:thresh}), we say that it is a \emph{highly confident} normal sample if it also satisfies $S_T(\bm{h}_r) < \tau_l$. Then, we use the new embedding $\bm{h}_r$ to recalculate the $d$ histograms used in Equation~\eqref{eqn:histogram}, thereby updating our outlier detection model. The threshold value $\tau_l$ is again considered as a hyperparameter. As shall be shown in the next section, we empirically confirm that this online model update scheme indeed improves the in-out detection performance.

It is worth noting that the concept of leveraging the `predicted' labels of originally unlabeled samples to enhance a model has been around in the literature~\cite{lee2013pseudo,xie2020self}. Such predicted labels are often called \emph{pseudo labels}. It was first shown in~\cite{lee2013pseudo} that for the data samples naturally belonging to different clusters, the pseudo labels of the (originally unlabeled) samples can improve the model performance substantially.

\subsection{Summary and Complexity Analysis}
\label{subsec:algo}
We summarize the entire inference process in Algorithm~\ref{algo:inference} and provide its complexity analysis as follows. Observe that Line 1 is a $O(1)$ operation since a new signal record contains a limited number of MACs (normally less than 100 MACs). Also, Line 2 has the complexity of $O\left(K(dN_s \!+\! d^2)\right)$ as can be seen from the complexity analysis of Algorithm~\ref{algo:bisage}, where $d$ is the embedding dimension. Line 3 requires $O(d)$ operations as the outlier score $S_T(\bm{h}_r)$ is computed along each embedding dimension. In addition, the complexity of Lines 4--8 is dominated by the model-update operation in Line 7 which has the complexity of $O(d|U|)$. It is because $d$ histograms are recalculated based on all in-boundary signal records, including the new highly confident signal record, and the record size is bounded by $|U|$. Note that only a small portion of new records would be used to update the model in practice. Therefore, the time complexity of Algorithm~\ref{algo:inference} is $O\left(K(dN_s \!+\! d^2) \!+\! d|U| \right)$.

\section{Performance Evaluation}
\label{sec:exp}
In this section, we present extensive experiment results to demonstrate the efficacy of \n{}. We evaluate its system-level performance by comparing it with state-of-the-art algorithms. We also empirically demonstrate the effectiveness of each system component.

\vspace{1mm}
\noindent\textbf{Experiment setup:} We developed an Android application to collect RSS values from ambient APs and notify if the user is out of the area. The application continuously collects RF signal records and uploads them to a server that performs in-out detection based on the gathered data.

We recruited 10 volunteers in the experiments. A user carries a smartphone such as Samsung S7, Huawei Nova 6 or Xiaomi Mix 3, or a smartwatch, i.e., alps DM20, in a typical home setting, e.g., a single-room dorm, a small or large apartment, or a two-story house. The housing area ranges from $10~\text{m}^2$ to $200~\text{m}^2$. Each user is asked to walk around the \emph{inner} perimeter of the house for just a few minutes (i.e., 5-10 minutes) for initial training. Then, the user can behave as he/she wishes, i.e., staying inside or moving outside of the house for testing. The whole process lasts for about three hours for each user. To further evaluate the performance of \n{} under different environmental factors, we also test \n{} with the signal records collected for three days around our lab. In addition, we use other large datasets to study the scalability of \n{}.

We use the following baseline parameters for \n{}, which were obtained through hyperparameter tuning. The learning rate is 0.003, the embedding dimension is 32, the offset $c$ is 120 dBm, the scaling factor $T$ is 0.06, the in-out threshold $\tau_u$ is 0.005, and the updating threshold $\tau_l$ is 0.001. All experiments were conducted on a server that operates on Ubuntu 18.04 and has an Intel Core i9-9900X with 10 cores at 3.5~GHz, 64-GB memory, and an Nvidia 2080Ti GPU.

\vspace{2pt}
\noindent\textbf{Geofencing algorithms for performance comparison:} We consider the following state-of-the-art algorithms for performance evaluation. By noting that \n{} consists of \bisage{} and our outlier detection algorithm (dubbed `OD'), we also consider other network embedding and outlier detection algorithms, each of which is used together with OD or \bisage{}.

\begin{table*}[t]
\caption{Performance comparison with state-of-the-art algorithms. The entry is in the form of mean (min, max). OD indicates our histogram-based outlier detection algorithm, and \bisage{} indicates our network embedding algorithm.}
\vspace{-0.1in}
\centering 
\resizebox{\textwidth}{!}{
\begin{tabular}{l c c c c c c c c c c}
\toprule
 \multirow{2}{*}{Algorithms}&\hspace{0.1in}& \multicolumn{3}{c}{In-premises detection} &\hspace{0.1in}& \multicolumn{3}{c}{Outside detection} \\
 && $P_{in}$ & $R_{in}$ & $F_{in}$ & \hspace{0.1in} & $P_{out}$ & $R_{out}$ & $F_{out}$ \\
\toprule
\n{} (\bisage{} + OD) && \textbf{0.98} (0.94, 1.00)  & \textbf{0.99} (0.94, 1.00) & \textbf{0.98} (0.97, 1.00) &\hspace{0.1in}& \textbf{0.98} (0.88, 1.00) & \textbf{0.97} (0.88, 1.00) & \textbf{0.97} (0.94, 1.00)\\
\hline
SignatureHome && 0.98 (0.86, 1.00) & 0.97 (0.92, 1.00) & 0.97 (0.88, 0.98) &\hspace{0.1in}& 0.80 (0.62, 0.90) & 0.86 (0.76, 0.95) & 0.83 (0.70, 0.93)\\
\hline
INOA && 0.95 (0.75, 0.98) & 0.82 (0.57, 0.88) & 0.88 (0.65, 0.93) &\hspace{0.1in}& 0.69 (0.50, 0.85) & 0.91 (0.67, 0.98) & 0.78 (0.56, 0.90)\\
\hline
\rule{0pt}{3ex}
Other algorithms when integrated with \bisage{} or OD
\vspace{0.05in}
\\
\hline
GraphSAGE + OD && 0.82 (0.70, 0.94) & 0.97 (0.78, 0.99) & 0.88 (0.75, 0.95) &\hspace{0.1in}& 0.85 (0.65, 0.92) & 0.70 (0.61, 0.85) & 0.78 (0.64, 0.88)\\
\hline
Autoencoder + OD && 0.97 (0.79, 1.00) & 0.88 (0.67, 0.92) & 0.92 (0.72, 0.95) &\hspace{0.1in}& 0.52 (0.27, 0.70) & 0.82 (0.61, 0.91) & 0.64 (0.41, 0.78)\\
\hline
MDS + OD && 0.98 (0.82, 1.00) & 0.83 (0.63, 0.94) & 0.90 (0.73, 0.93) &\hspace{0.1in}& 0.43 (0.15, 0.68) & 0.87 (0.62, 0.96) & 0.58 (0.32, 0.72)\\
\hline
\bisage{} + Feature bagging &&  0.91 (0.78, 0.95) & 0.85 (0.78, 0.93) & 0.87 (0.80, 0.93) &\hspace{0.1in}& 0.85 (0.72, 0.90) & 0.93 (0.84, 0.98) & 0.89 (0.79, 0.94)\\
\hline
\bisage{} + iForeast && 0.86 (0.70, 0.91) & 0.97 (0.89, 1.00) & 0.90 (0.79, 0.93) &\hspace{0.1in}& 0.94 (0.86, 1.00) & 0.80 (0.68, 0.87) & 0.86 (0.78, 0.91)\\
\hline
\bisage{} + LOF && 0.89 (0.75, 0.95) & 0.89 (0.76, 0.96) & 0.89 (0.74, 0.95) &\hspace{0.1in}& 0.88 (0.72, 0.90) & 0.83 (0.68, 0.92) & 0.85 (0.72, 0.90)\\
\bottomrule
\end{tabular}
}
\label{tab:results_all}
\vspace{-0.2in}
\end{table*}

\begin{itemize}[itemsep=1pt]
    \item SignatureHome~\cite{SignatureHome}: It builds a `home signature' database for the geofencing area, containing the union of the detected MACs from all RF signal records collected inside the area and the IP address of the user's associated AP. When it comes to online inference, for each newly collected signal record, it checks whether the connected IP exists in the home signature (assigned a weight) and calculates the overlap ratio of MACs between the new record and the home signature (assigned another weight). If the total weight is higher than a threshold, the record is inside the area.    
    
    \item INOA~\cite{INOA18}: It trains an ensemble of base learners where each base learner learns a hypersphere based on RF signal records from each pair of APs. For inference, a new RF signal record is decomposed into corresponding pairs of APs, which are then fed into the base learners to obtain an outlier score. If it is higher than a threshold, the record is outside the area.
    
    \item GraphSAGE~\cite{graphsage} + OD: GraphSAGE replaces \bisage{} in \n{} to obtain node embeddings from our weighted bipartite graph by treating the graph as a \emph{homogeneous} graph. We then use the learned embeddings to build an outlier detection model as illustrated in Section~\ref{subsec:boundary_training}. For inference, upon the arrival of a new RF signal record, we first obtain its embedding via GraphSAGE and then perform outlier detection and model update as explained in Section~\ref{subsec:inout_prediction} and Section~\ref{subsec:b_update}, respectively.
    
    \item Autoencoder~\cite{Goodfellow-et-al-2016} + OD: Autoencoder uses an encoding and decoding process to learn (low-dimensional) embeddings from a matrix of signal records, where the missing entries are filled with small values. The embeddings generated by the autoencoder replace those generated by \bisage{} and the rest follows as in GraphSAGE + OD.
    
    \item Multidimensional scaling (MDS)~\cite{cox2008multidimensional} + OD: MDS learns the embeddings from a matrix of pairwise distances calculated from signal records, where missing entries are padded with small values. The embeddings generated by MDS replace those generated by \bisage{} and the rest follows as in GraphSAGE + OD.
    
    \item \bisage{} + feature bagging~\cite{kdd05FeatureBagging}: Feature bagging designs a framework to fuse the results from different outlier detection algorithms based on different subsets of features, which are the embeddings generated by \bisage{}. When a new RF signal record is available for inference, its embedding is first learned by \bisage{} and then goes through the feature bagging framework to decide whether it is an outlier.
    
    \item \bisage{} + isolation forest (iForest)~\cite{IsolationForest}: iForest finds outliers with the rationale that outliers are easier to `isolate' from normal (in-premises) samples. It also uses the embeddings generated by \bisage{}. The rest follows as in \bisage{} + feature bagging.
    
    \item \bisage{} + local outlier factor (LOF)~\cite{lof}: LOF finds outliers by observing that the distance from an outlier to a neighbor group is larger than the distance inside the neighbor group. It also uses the embeddings generated by \bisage{}. The rest follows as in \bisage{} + feature bagging.
    
\end{itemize}

The algorithms with \bisage{} use the embeddings generated by \bisage{} as input and perform in-out detection. The algorithms with OD first generate the embeddings on their own and carry out in-out detection with our outlier detection algorithm. For SignatureHome and INOA, we set their parameters as described in~\cite{SignatureHome,INOA18}. For autoencoder, its best results, which are obtained using four layers of 1-D convolution with the ReLU activation function, are used for performance comparison. For MDS, we follow the convention with $1\!-\!\text{cosine similarity}$ as pairwise distance.

\vspace{2pt}
\noindent\textbf{Performance metrics:} We use precision $P$, recall $R$, and $F$-score to evaluate classification results in our experiments. Let $TP$, $FP$, and $FN$ be the number of true positives, the number of false positives, and the number of false negatives, respectively. Then, we have $P \!=\! \frac{TP}{TP + FP}$, $R \!=\! \frac{TP}{TP + FN}$, and $F \!=\! \frac{2PR}{P + R}$. To better understand the system performance, we use $P_{in}$, $R_{in}$ and $F_{in}$ for in-premises detection in which case in-premises samples are treated as positive samples, and $P_{out}$, $R_{out}$ and $F_{out}$ for outlier detection in which case outside samples are considered positive.

\subsection{Overall Performance}
We show overall comparison results of \n{} against other state-of-the-art algorithms in Table~\ref{tab:results_all}. \n{} achieves the \emph{best} performance in all the metrics. This is because the bipartite graph modeling of RF signal records and \bisage{} have well-learned similarities among signal records. Moreover, our outlier detection algorithm is also more robust to statistical fluctuations in outlier score prediction. Thus, outliers can be easily identified. SignatureHome has relatively low precision and recall in outside detection, indicating that it may have problems in separating signals observed near the boundary of the house since its network-based approach is not able to capture any perimeter information. INOA suffers from low precision in outside detection, which means that the support vectors generated from subsets of MACs do not represent `inside' signals well and in turn lead to lower recall for in-premises detection. Furthermore, as mentioned in Section~\ref{subsec:bipartite}, both SignatureHome and INOA have the missing-value problem, where a matrix representation of RF signals requires missing entries to be filled with some small values (i.e., $-120$ dBm), thereby degrading their performance.

\vspace{2pt}
\noindent\textbf{Effectiveness of \bisage{}:} To study the effectiveness of our bipartite graph modeling and \bisage{} in \n{}, we show the performance comparison of \n{}, GraphSAGE, autoencoder, and MDS in Table~\ref{tab:results_all}. For autoencoder and MDS, we construct a matrix of RF signal records, with missing entries filled with $-120$ dBm. As shown in Table~\ref{tab:results_all}, \n{} exhibits superior performance over the other algorithms (24\%--67\% in $F_{out}$). This verifies that the embeddings learned from \bisage{} are much more accurate in preserving the similarities and differences among signal records. On the other hand, GraphSAGE treats the bipartite graph as a homogeneous graph. It does not take into account node heterogeneity in the aggregation process, and its resulting embeddings are less accurate in representing the observed samples and their relationships. Autoencoder has low recall for in-premises detection, indicating that quite a few in-premises records are misclassified as outliers, i.e., $P_{out}$ is low. This implies that the missing-value problem that arises when using a matrix for RF signals results in inaccurate inference of the similarities among signal records. A similar issue happens with MDS.

\begin{figure}[t]
    \centering
	\subfloat[]{%
    \includegraphics[width=0.5\linewidth]{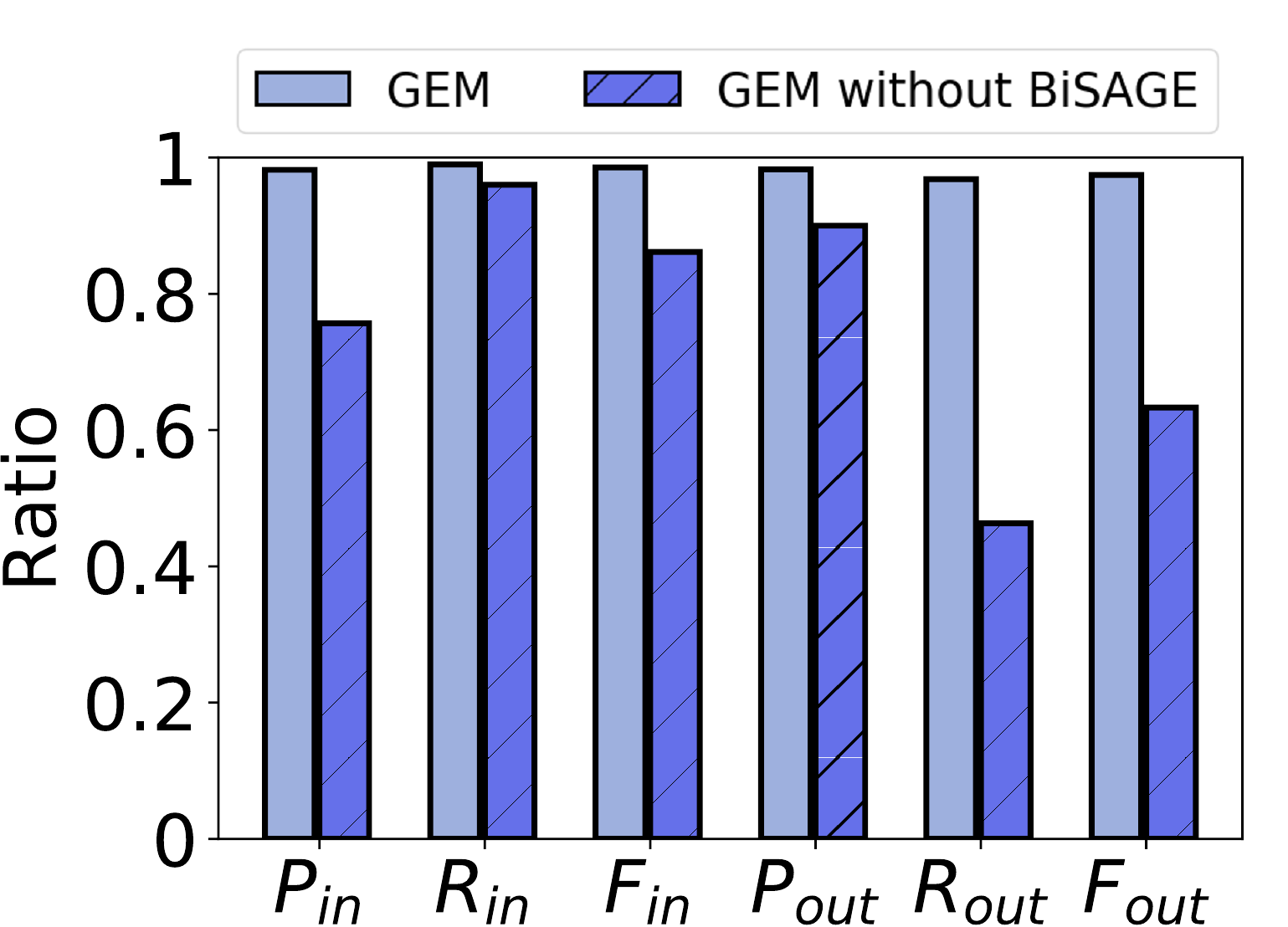}
    }
    \subfloat[]{%
    \includegraphics[width=0.45\linewidth]{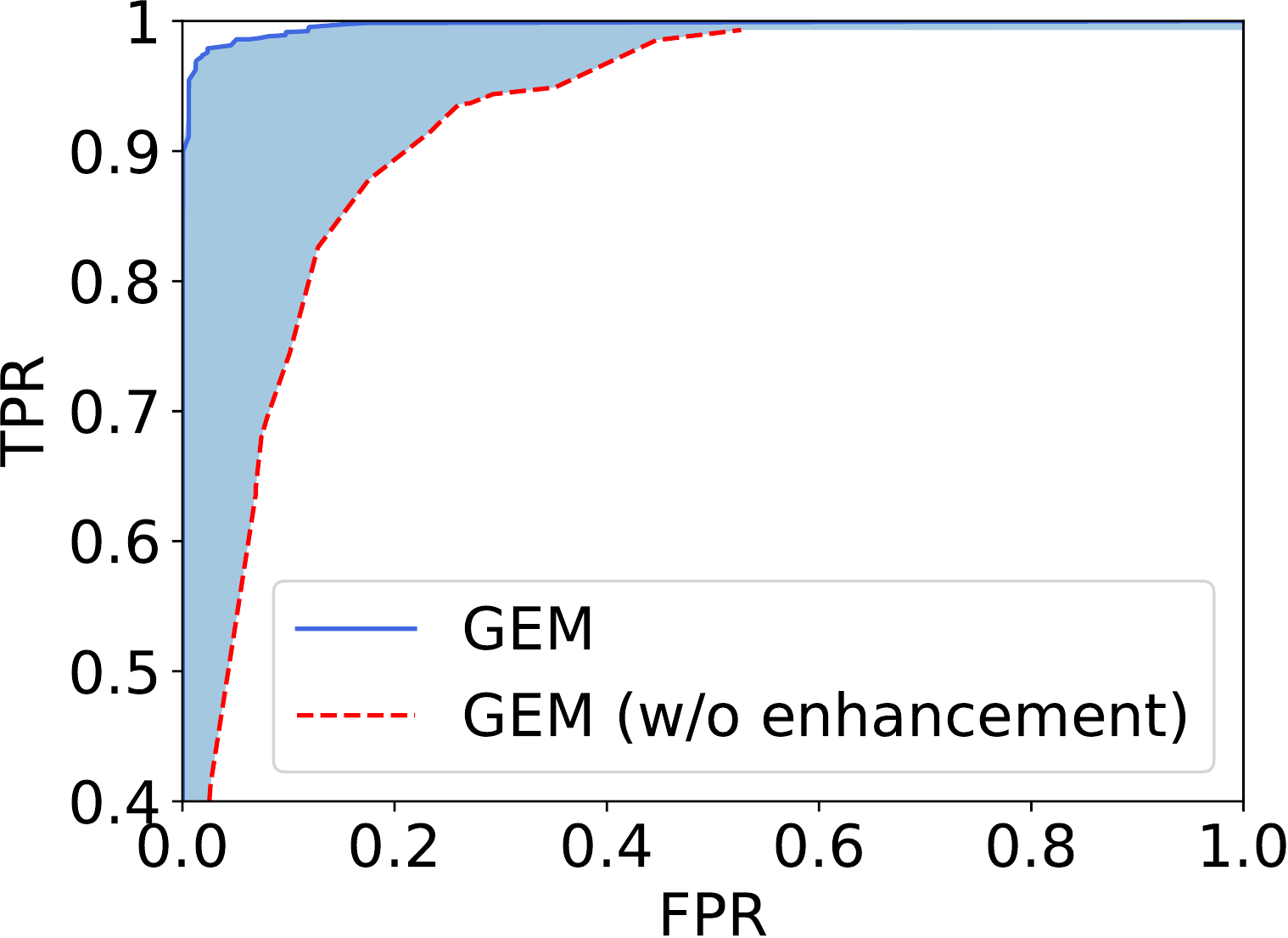}
    }
    \vspace{-0.05in}
	\caption{(a) Performance comparison of \n{}  and \n{} without the embeddings by \bisage{}; (b) Performance comparison of \n{} with and without our proposed enhancement in the histogram-based outlier detection algorithm.}
	\vspace{-0.2in}
	\label{fig:comparison_bisage}
\end{figure}

To see the gains in \n{} owing to the bipartite graph modeling and \bisage{}, we show the performance comparison results of \n{} with and without using the embeddings by \bisage{} in Figure~\ref{fig:comparison_bisage}(a). For \n{} without the embeddings, we construct a matrix of signal records and fill in empty entries with $-120$ dBm. As seen from Figure~\ref{fig:comparison_bisage}(a), \n{} improves the performance significantly (around 14\% in $F_{in}$ and 54\% in $F_{out}$) when using embeddings as input, since \bisage{} is able to effectively learn the similarities among RF signals.

\vspace{2pt}
\noindent\textbf{Effectiveness of our outlier detection algorithm:} We use the same embeddings learned from \bisage{} as input for different outlier detection algorithms, namely feature bagging, iForest, and LOF, and compare their performance with \n{} in Table~\ref{tab:results_all}. As can be seen from Table~\ref{tab:results_all}, \n{} has the highest $F$-scores in both in-premises detection and outside detection, with \emph{further} improvements of up to 12\% in $F_{in}$ and 18\% in $F_{out}$. This is because our enhanced outlier detection algorithm `stabilizes' statistical fluctuations and achieves better in-out classification performance along with new samples over time. iForest also has high recall for in-premises detection. However, its precision is not as good as that of \n{}, which indicates that iForest is not able to filter out some outside signals properly. LOF does not show satisfactory performance in outside detection as it may not be able to classify the signals around the house boundary. Feature bagging may encounter a similar issue in outlier detection.
 
To see the effectiveness of our enhancement in the histo-gram-based outlier detection, we plot the receiver operating characteristic (ROC) curves~\cite{zweig1993receiver} for \n{} with and without our enhancement in Figure~\ref{fig:comparison_bisage}(b).\footnote{The ROC curve plots true positive rate vs. false positive rate (or false alarm rate) with \emph{varying} threshold values, so it shows the overall performance of a classification model across different threshold values.} Here we focus on in-premises detection. For the ROC curve of a model, the closer to the upper left corner (a perfect classifier), i.e., the larger area under the ROC curve, the better the model is. As shown in Figure~\ref{fig:comparison_bisage}(b), with the same true positive rate, \n{} achieves a much lower false-positive rate, implying that it greatly reduces the chance of missing the detection of outliers. We also observe that recall of outlier detection can \emph{hardly} be improved by the \emph{original} histogram-based outlier detection algorithm. This indicates that a non-trivial portion of the signal records obtained outside the geofencing area are consistently misclassified. On the contrary, our enhancement significantly improves the outlier detection performance by clearly separating the two groups of normal and abnormal signal records.

\begin{figure}[t]
    \centering
	\subfloat[]{%
    \includegraphics[width=0.48\linewidth]{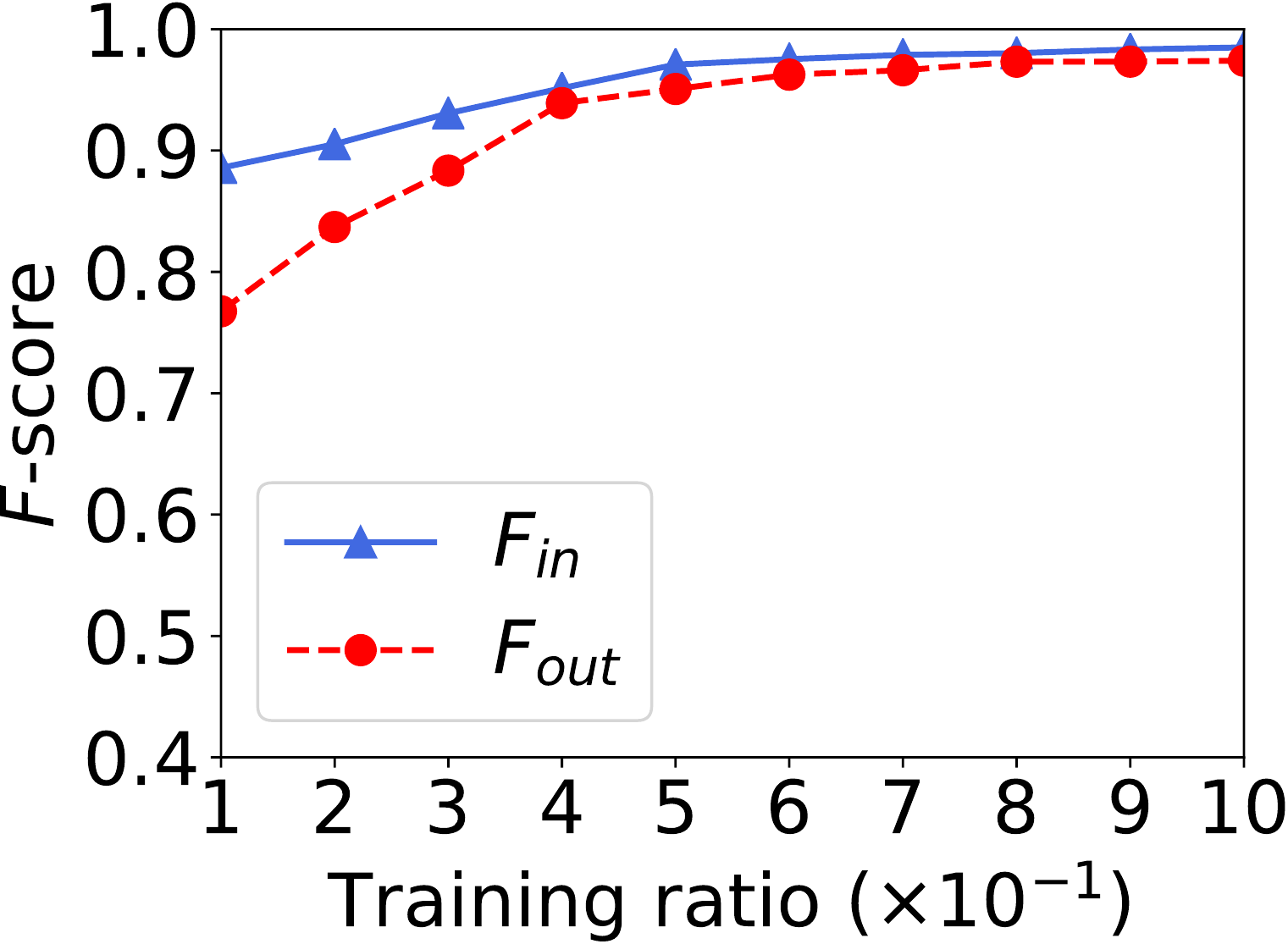}
    }
    \subfloat[]{%
    \includegraphics[width=0.48\linewidth]{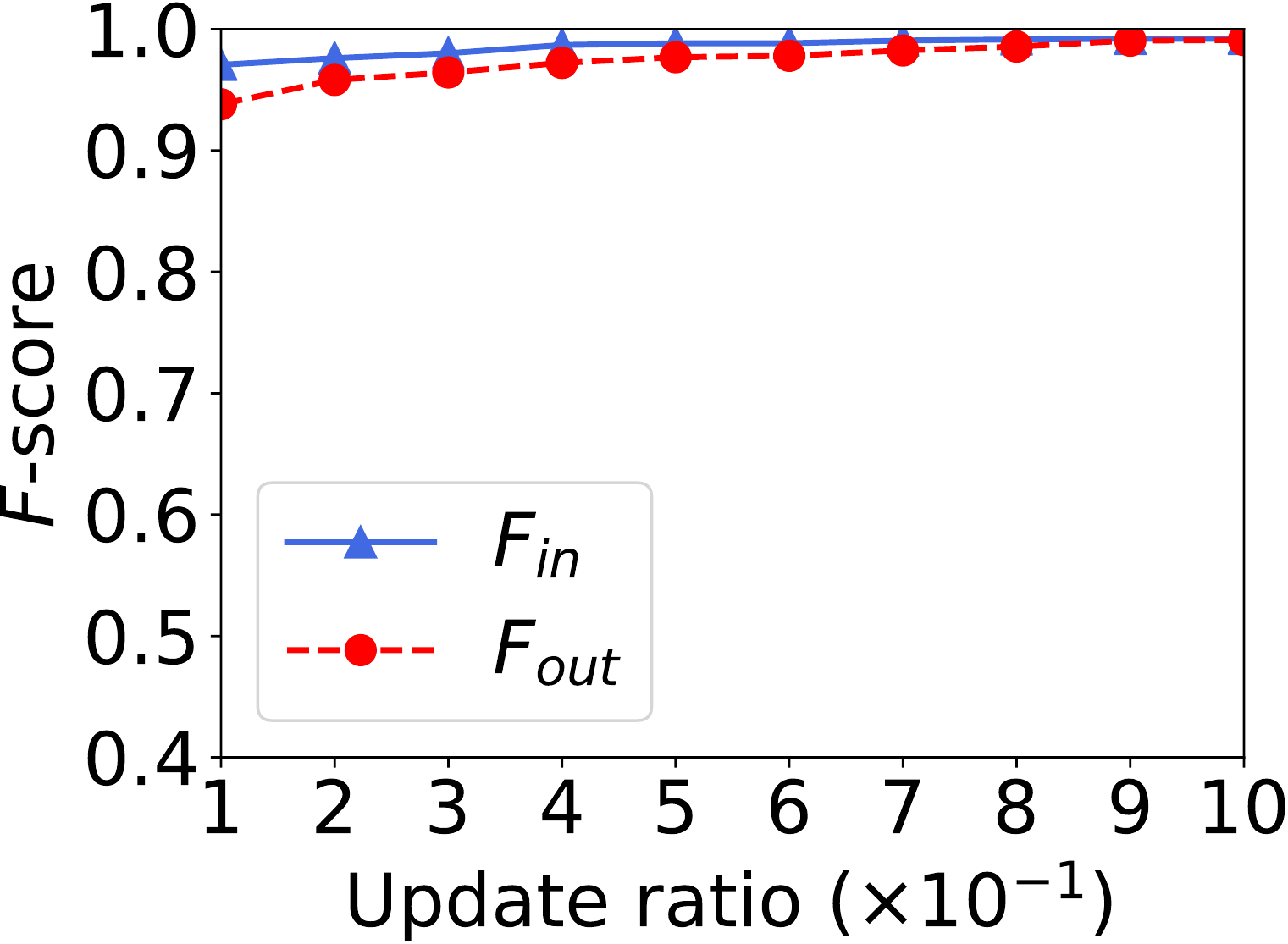}
    }
    \vspace{-0.05in}
	\caption{(a) Performance vs. training ratio; (b) Performance vs. update ratio.}
	\vspace{-0.1in}
	\label{fig:model_update}
\end{figure}

\vspace{2pt}
\noindent \textbf{Benefits of model training and its online update:} To see how many samples are needed for training our outlier detection model, we divide the (initial) training samples into ten equal sets and train \n{} with one more set each time to perform outlier detection. As shown in Figure~\ref{fig:model_update}(a), the performance gradually improves with more training samples available, and \n{} can \emph{even} work with only 10\% of training samples (less than 50 records on average), indicating its practicability in real deployment. A key design of \n{} is that it leverages new incoming samples to improve the in-out classification accuracy over time. To evaluate this self-enhancement, we divide the `testing' data into ten equal sets and present step-wise model update results in Figure~\ref{fig:model_update}(b). The performance of \n{} improves with new samples streaming in, showing that it leverages highly confident in-premises samples to improve the detection accuracy.

\begin{table}[t]
\caption{User-level performance of \n{}.}
\vspace{-0.1in}
\centering 
\resizebox{0.47\textwidth}{!}{
\begin{tabular}{c c c c c c c c c}
\toprule
 \multirow{2}{*}{User}& \multicolumn{3}{c}{In-premises detection} & \multicolumn{3}{c}{Outside detection} & \multirow{2}{*}{\#MACs} & Area \\
 & $P_{in}$ & $R_{in}$ & \multicolumn{1}{c}{$F_{in}$} & $P_{out}$ & $R_{out}$ & $F_{out}$ && ($m^2$)\\
\toprule
1 & 1.00 & 0.94 & 0.97 & 0.97 & 1.00 & 0.99 & 20 & $\sim\!10$\\
\rowcolor{Gray}
2 & 0.94 & 1.00 & 0.97 & 1.00 & 0.97 & 0.98 & 26 & $\sim\!10$\\
3 & 1.00 & 0.99 & 0.99 & 0.99 & 1.00 & 0.99 & 33 & $\sim\!50$\\
\rowcolor{Gray}
4 & 0.98 & 1.00 & 0.99 & 1.00 & 0.88 & 0.94 & 16 & $\sim\!50$\\
5 & 0.96 & 1.00 & 0.98 & 1.00 & 0.88 & 0.94 & 20 & $\sim\!50$\\
\rowcolor{Gray}
6 & 0.98 & 1.00 & 0.99 & 1.00 & 0.94 & 0.97 & 65 & $\sim\!100$\\
7 & 1.00 & 0.97 & 0.98 & 0.88 & 1.00 & 0.94 & 45 & $\sim\!100$\\
\rowcolor{Gray}
8 & 1.00 & 1.00 & 1.00 & 1.00 & 1.00 & 1.00 & 73 & $\sim\!100$\\
9 & 1.00 & 0.98 & 0.97 & 1.00 & 0.98 & 0.99 & 57 & $\sim\!100$\\
\rowcolor{Gray}
10 & 0.94 & 1.00 & 0.97 & 1.00 & 0.98 & 0.99 & 12 & $\sim\!200$\\
\bottomrule
Avg. & 0.981 & 0.989 & 0.985 & 0.982 & 0.968 & 0.974 & 36.7 & $\sim\!77$\\
\bottomrule
\end{tabular}
}
\label{tab:results_ind}
\end{table}

\subsection{User-level Performance}
\label{sec:user}

We show quantitative results for each user in Table~\ref{tab:results_ind}. As seen from Table~\ref{tab:results_ind}, \n{} performs well (with most $F$-scores higher than $0.95$) across various housing types with different numbers of MACs seen. For multiplex apartments ($\leq 100~\text{m}^2$), more MACs lead to higher accuracy. On the other hand, for a two-story house ($\sim\!200\text{m}^2$), which is `detached' from others, although the number of MACs is small, it should be easier to detect outside events. Thus, \n{} remains effective for accurate geofencing. All the user studies validate the practicability of \n{}.

\subsection{Micro-benchmark Evaluation}

\label{subsec:exp_micro}
\begin{figure}[t]
\vspace{-0.15in}
    \begin{minipage}{0.49\textwidth}
        \centering
    	\subfloat[In-premises detection]{%
        \includegraphics[width=0.49\linewidth]{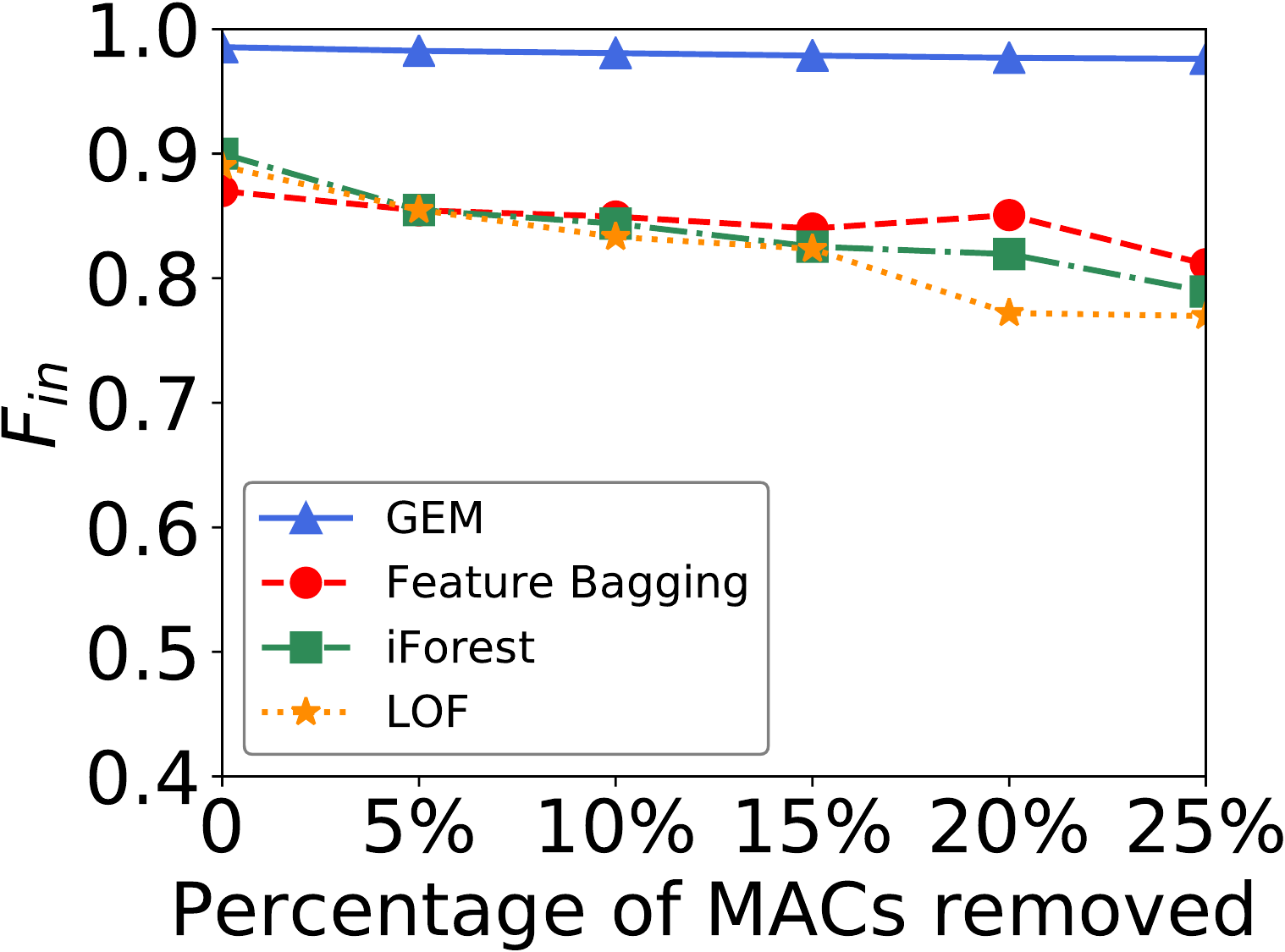}
        }
        \subfloat[Outside detection]{%
        \includegraphics[width=0.49\linewidth]{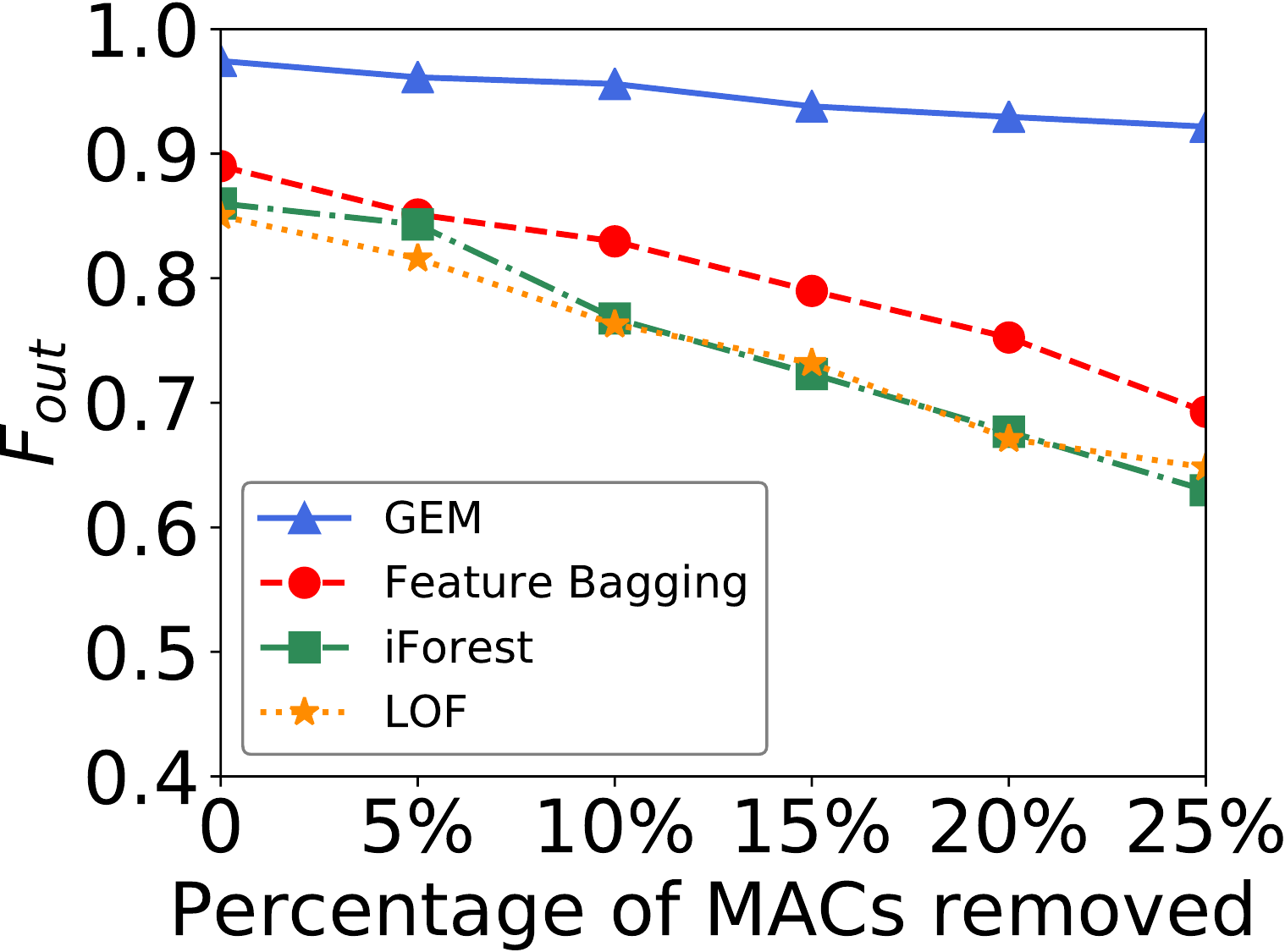}
        }
        \vspace{-0.02in}
    	\caption{Performance vs. MAC removal ratio in training set.}
    	\vspace{-0.2in}
    	\label{fig:prune_train}
	\end{minipage}
\end{figure}

\begin{figure}[t]
	\begin{minipage}{0.49\textwidth}
        \centering
    	\subfloat[In-premises detection]{%
        \includegraphics[width=0.49\textwidth]{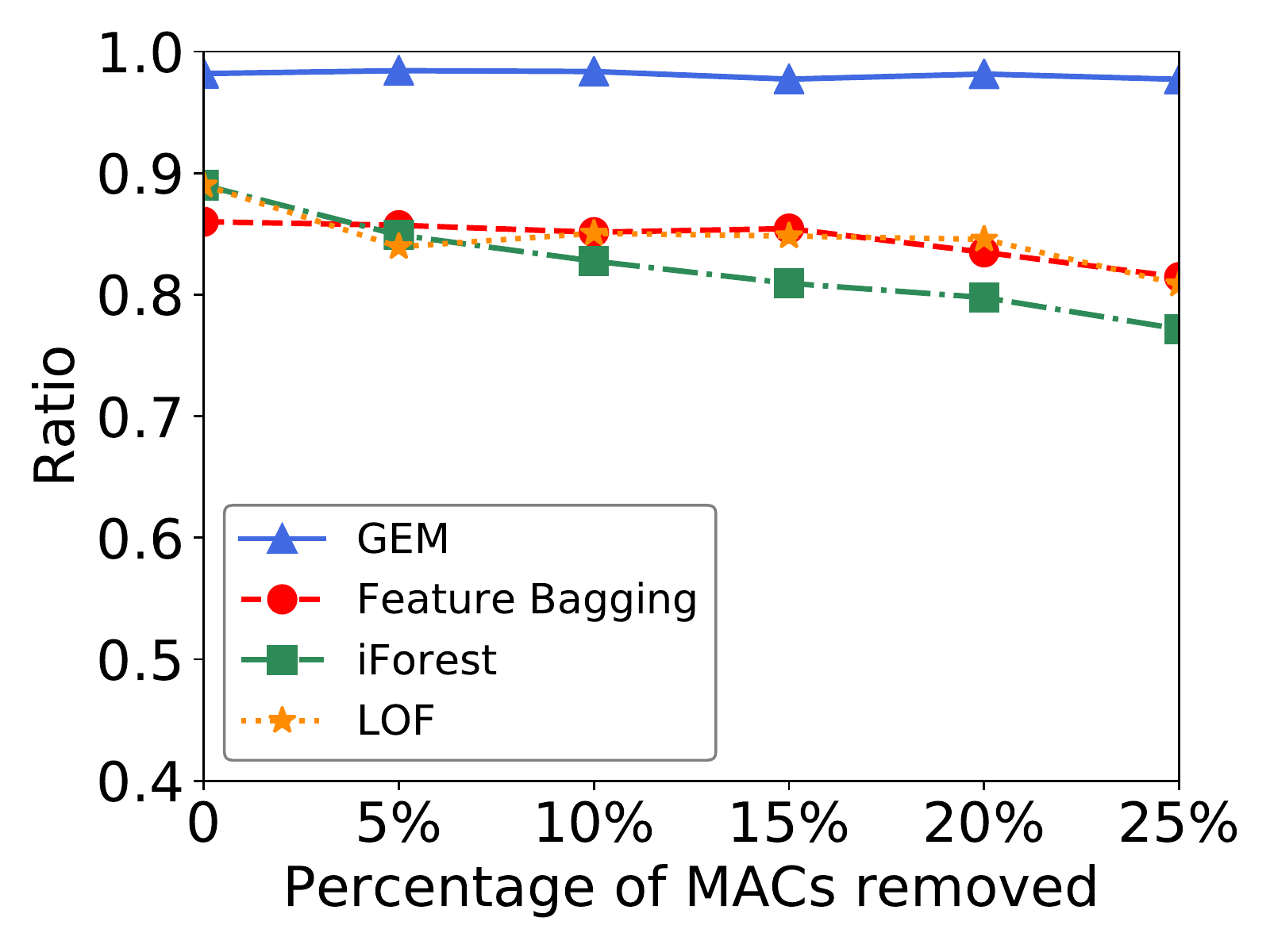}
        }
        \subfloat[Outside detection]{%
        \includegraphics[width=0.49\textwidth]{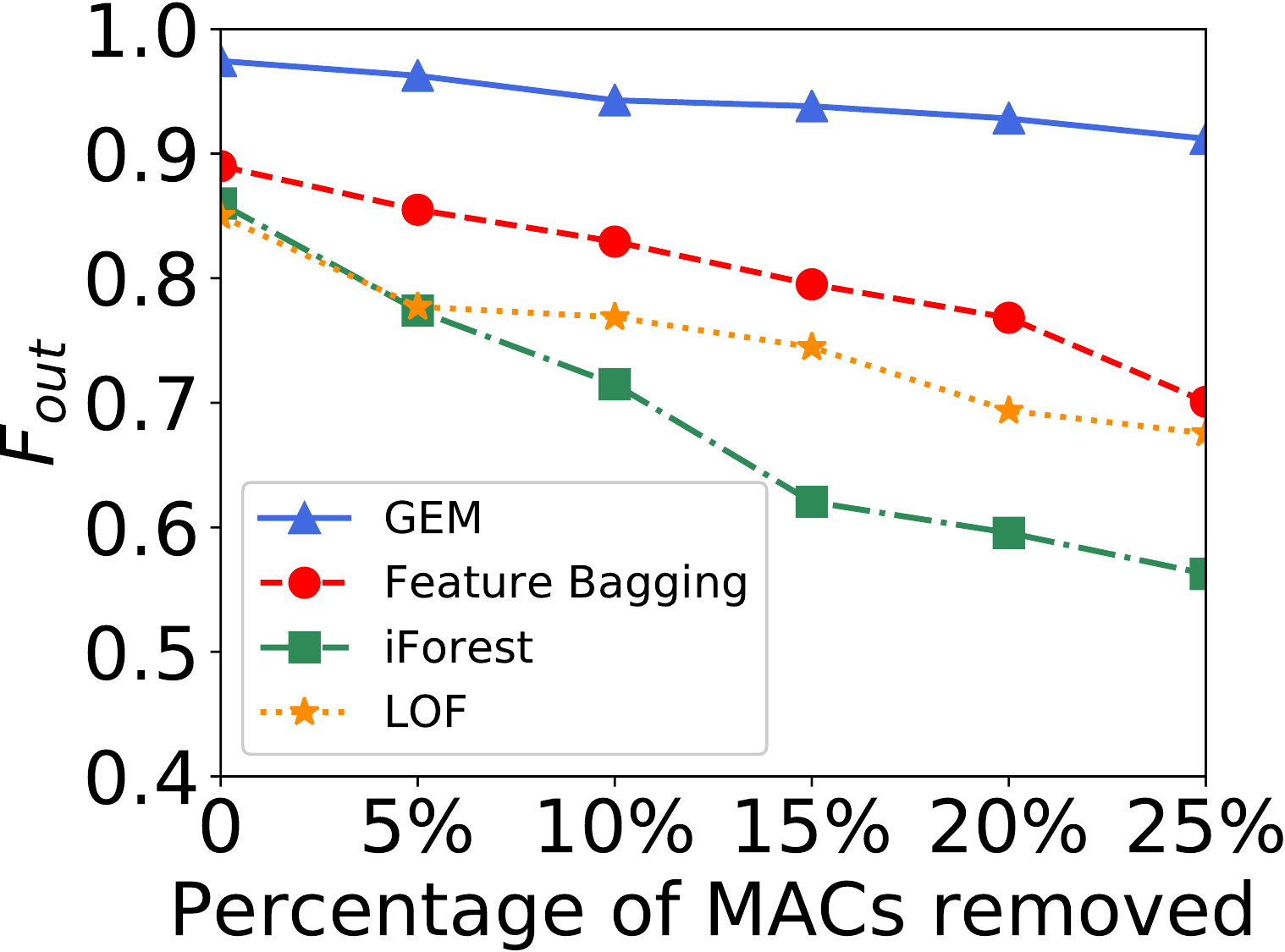}
        }
        \vspace{-0.02in}
    	\caption{Performance vs. MAC removal ratio in test set.}
    	\label{fig:prune_update}
	\end{minipage}
\end{figure}

\noindent\textbf{Adaptation to the changes in APs:} To test \n{}'s adaptability to AP dynamics, we intentionally remove a subset of APs from the training and testing sets. Note that they are all ambient APs sensed, and we do not install any additional AP on site. In each experiment, we randomly remove up to 25\% of MACs in total and repeat 30 runs to measure the average performance. Figure~\ref{fig:prune_train} shows $F$-scores by pruning MACs in the training set while leaving the testing set untouched. While $F$-scores of \n{} decrease slowly as more MACs are removed, it remains significantly better than the other algorithms. As shown in Figure~\ref{fig:prune_train}, the performance of \n{} for in-premises detection almost remains intact, while the performance for outside detection decreases only slightly. It is because \n{} keeps updating the outlier detection model with the highly confident in-premises samples. In other words, any newly sensed MACs can naturally be added into our bipartite graph, and they actually improve the in-out detection performance over time. Therefore, the removal of MACs in the training set does not significantly affect the performance. In addition, we intentionally remove MACs in the testing set, while leaving the training set unchanged. Similar results are observed, as shown in Figure~\ref{fig:prune_update}.

\begin{figure}[t]
	\centering
	\includegraphics[width=0.37\textwidth]{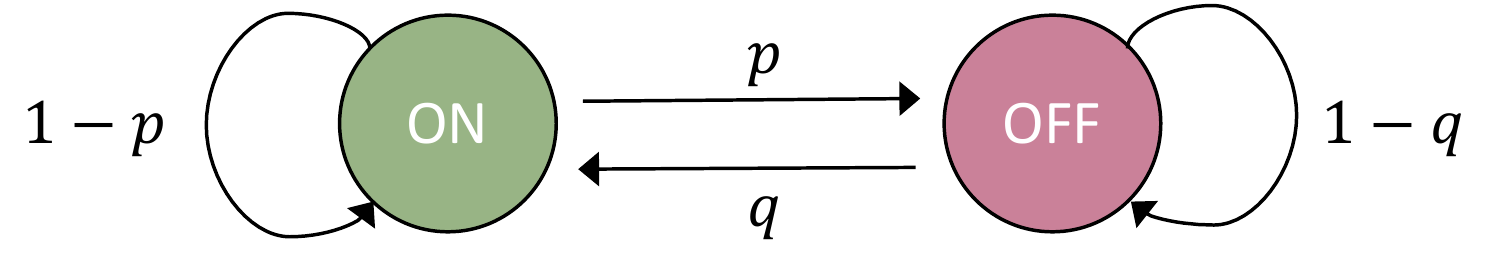}
	\vspace{-0.1in}
	\caption{Two-state Markov model for APs.}
	\label{fig:markov_state}
\end{figure}

\begin{figure}[t]
	\centering
	\includegraphics[width=0.35\textwidth]{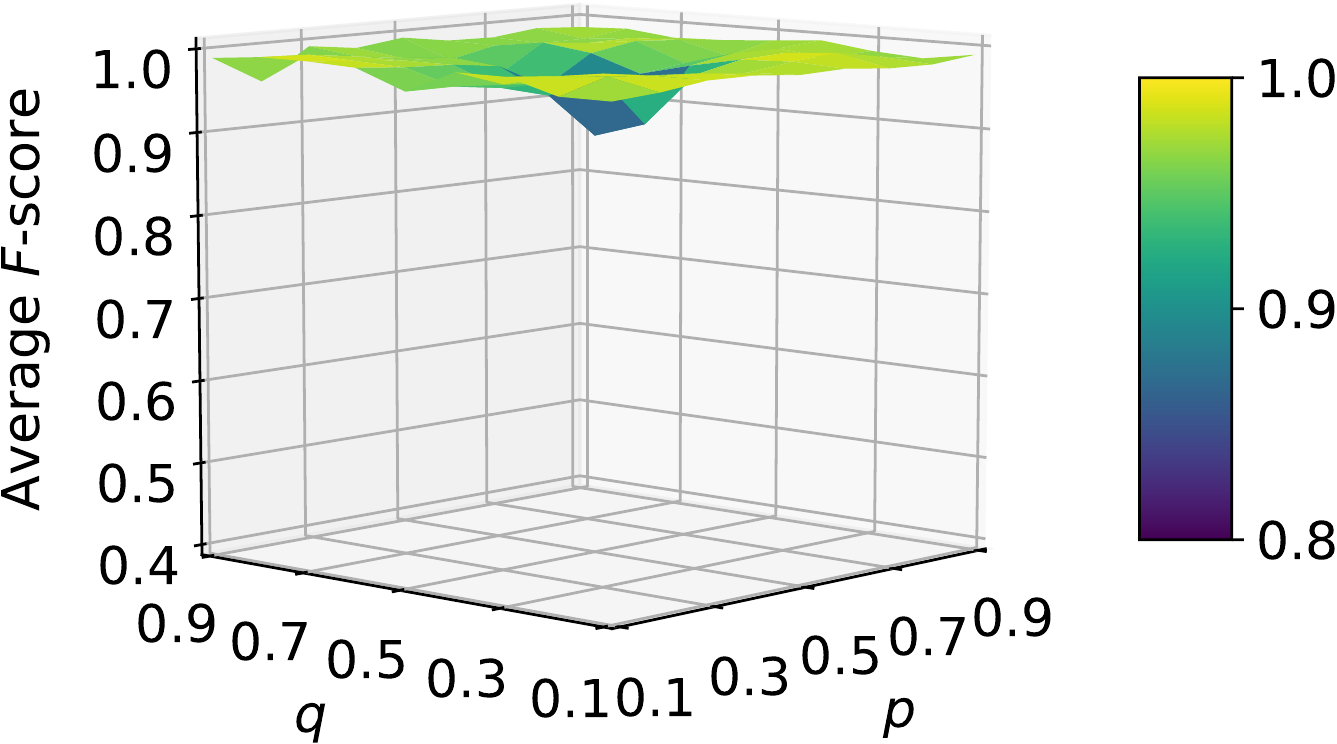}
	\vspace{-0.1in}
	\caption{Performance of \n{} is robust to signal dynamics.}
	\vspace{-0.2in}
	\label{fig:markov_accuracy}
\end{figure}

\noindent\textbf{Robustness to RF dynamics:} To further evaluate the robustness of \n{} to dynamic RF environments, we conduct an `AP ON-OFF' experiment where some RF records associated with each AP/MAC remain intact (`ON') or disappear (`OFF') in the training and testing sets, and the (dis)appearance is governed by a two-state Markov model, as depicted in Figure~\ref{fig:markov_state}. For each $(p, q)$ pair, each AP/MAC stays in its current state with probability $1 \!-\! p$ (resp. $1\!-\!q$) or moves to the other state with $p$ (resp. $q$), and each state transition, including self-transition, takes place every 30 samples throughout the training and testing sets. This experiment is repeated 30 times for in-premise and outside detections, and the average $F$ score is reported in Figure~\ref{fig:markov_accuracy}. Here we vary the values of $p$ and $q$ from 0.1 to 0.9. As shown in Figure~\ref{fig:markov_accuracy}, \n{} performs well regardless of the choices of $p$ and $q$, demonstrating its robustness to dynamic RF environments. It is worth noting that when $(p, q)$ gets close to (0.5, 0.5), the \emph{entropy rate} of its corresponding two-state Markov model increases, i.e., the Markov model has \emph{higher uncertainty}. The higher uncertainty of the model contributes to a small dip in the average $F$ score around $(p, q) \!=\! (0.5, 0.5)$. See~\cite{thomas2006elements} for more details on the entropy rate of Markov chains.

\begin{figure}[t]
    \vspace{-2mm}
    \centering        
    \subfloat[]{%
        \includegraphics[width=0.162\textwidth]{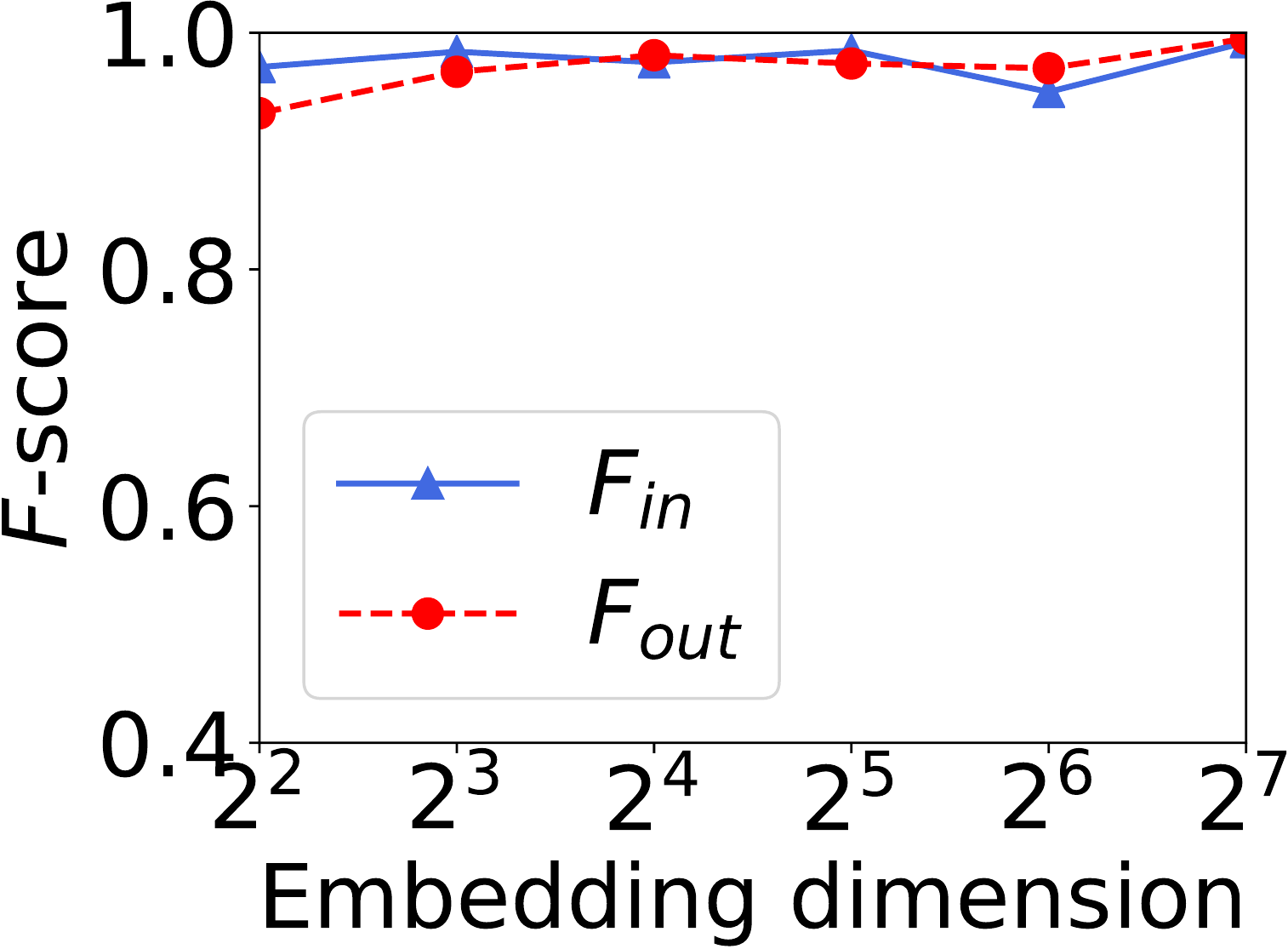}  }    
    \subfloat[]{%
        \includegraphics[width=0.162\textwidth]{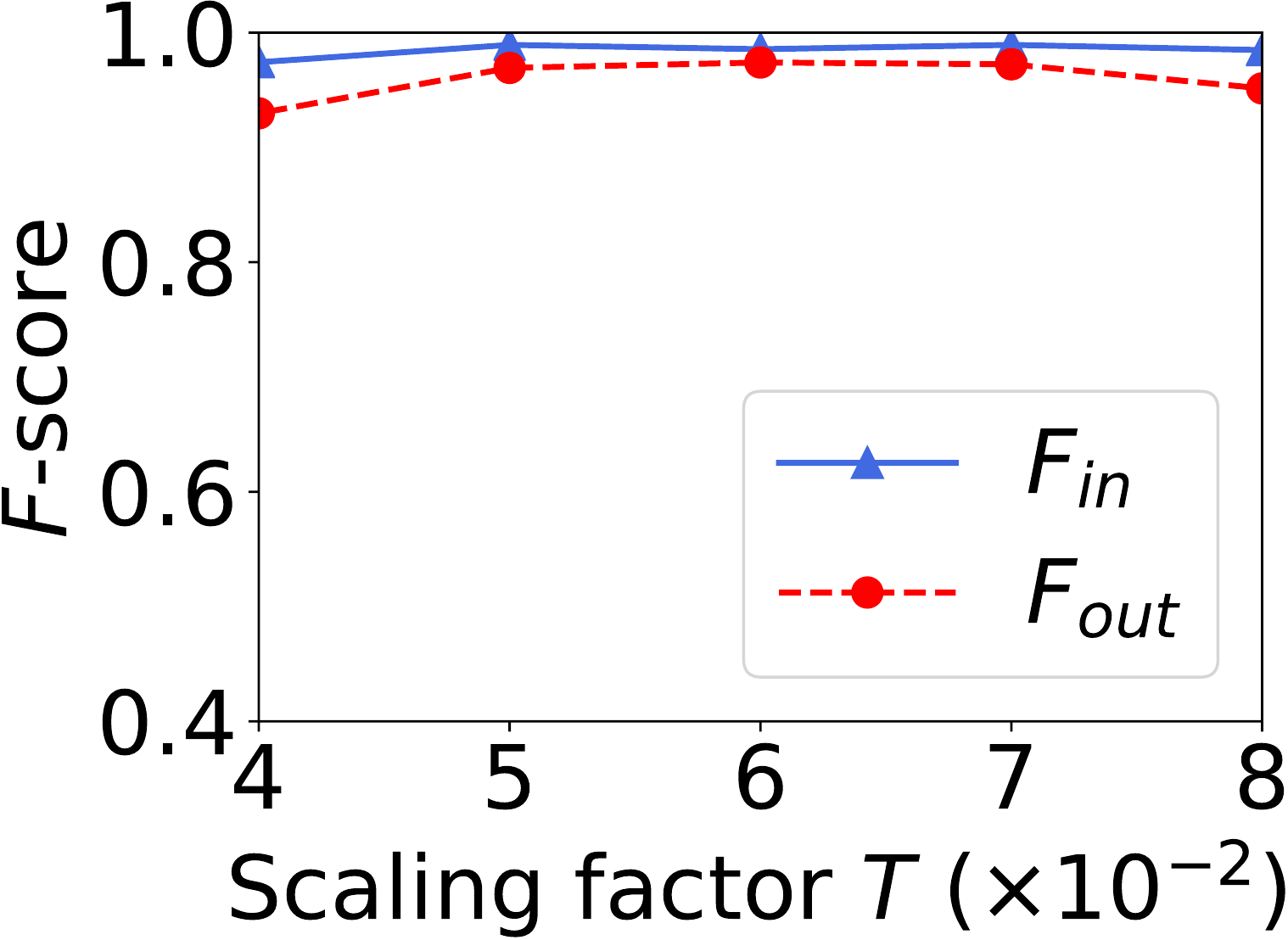}
        }    
    \subfloat[]{%
        \includegraphics[width=0.162\textwidth]{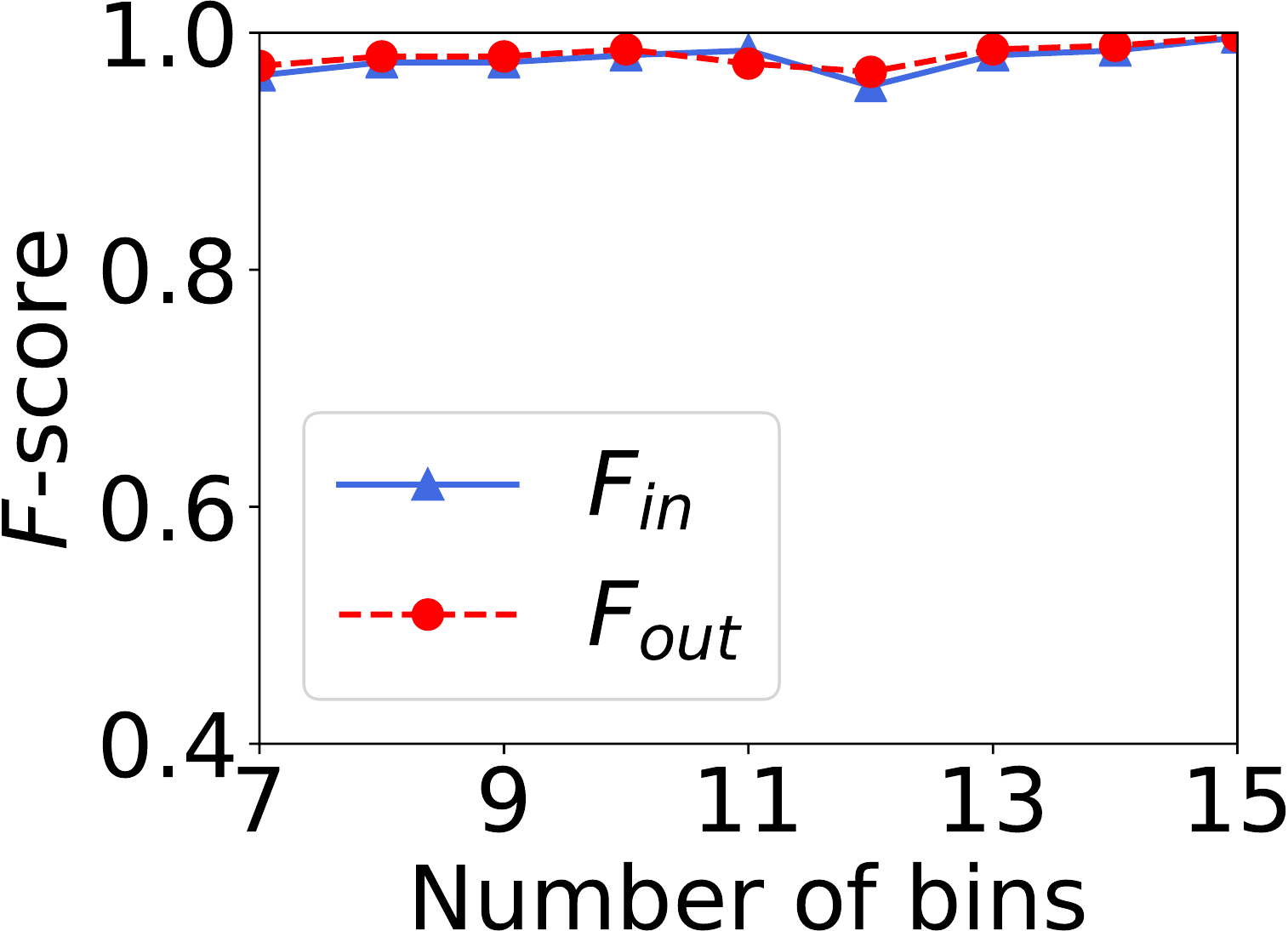}
        }    
    \vspace{-0.05in}
	\caption{Impact of (a) embedding dimension $d$, (b) scaling factor $T$, (c) bin size $m$, on the performance of \n{}.}
	\label{fig:robustness}
	\vspace{-0.2in}
\end{figure}

\vspace{2pt}
\noindent\textbf{Tolerance to parameter perturbation:} From the bipartite graph, \n{} learns node embeddings via \bisage{} and uses the embeddings for outlier detection. Thus, it is important to check how the choice of embedding dimension would affect the system performance. In Figure~\ref{fig:robustness}(a), we show the $F$-scores when the embedding dimension varies. We can see that \n{} exhibits outstanding performance regardless of the choices of the dimension, implying its robustness to such changes.

Similarly, in Figure~\ref{fig:robustness}(b), we present the overall system performance of \n{} when the scaling factor $T$ in Equation~\eqref{eqn:enhanced_score} changes. We can see that \n{} is not sensitive to the choices of $T$ from 0.04 to 0.08 and achieves very high $F$ scores overall. Thus, we can easily tune the scaling factor $T$ to achieve highly accurate performance in practice. In addition, Figure~\ref{fig:robustness}(c) presents the effect of the number of bins $m$ on the performance in outlier detection. Recall that our enhanced histogram-based outlier detection algorithm divides the range of values along each dimension into $m$ bins and builds $d$ histograms for outlier detection. Our algorithm achieves excellent performance over a wide range of values for $m$.

\begin{figure*}[t]
    \subfloat[]{%
    \includegraphics[width=0.2\linewidth]{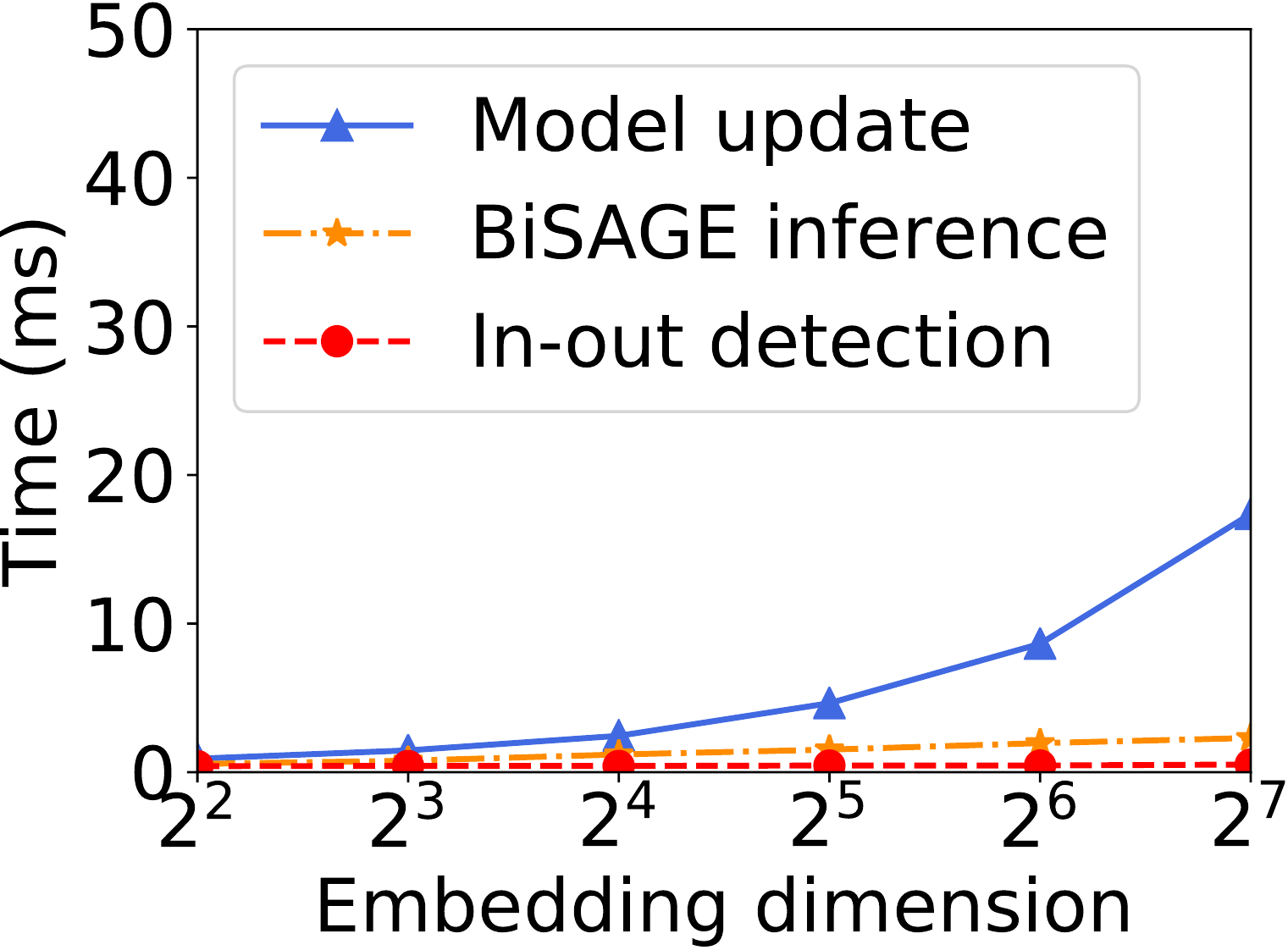}
    }
    \subfloat[]{%
    \includegraphics[width=0.2\linewidth]{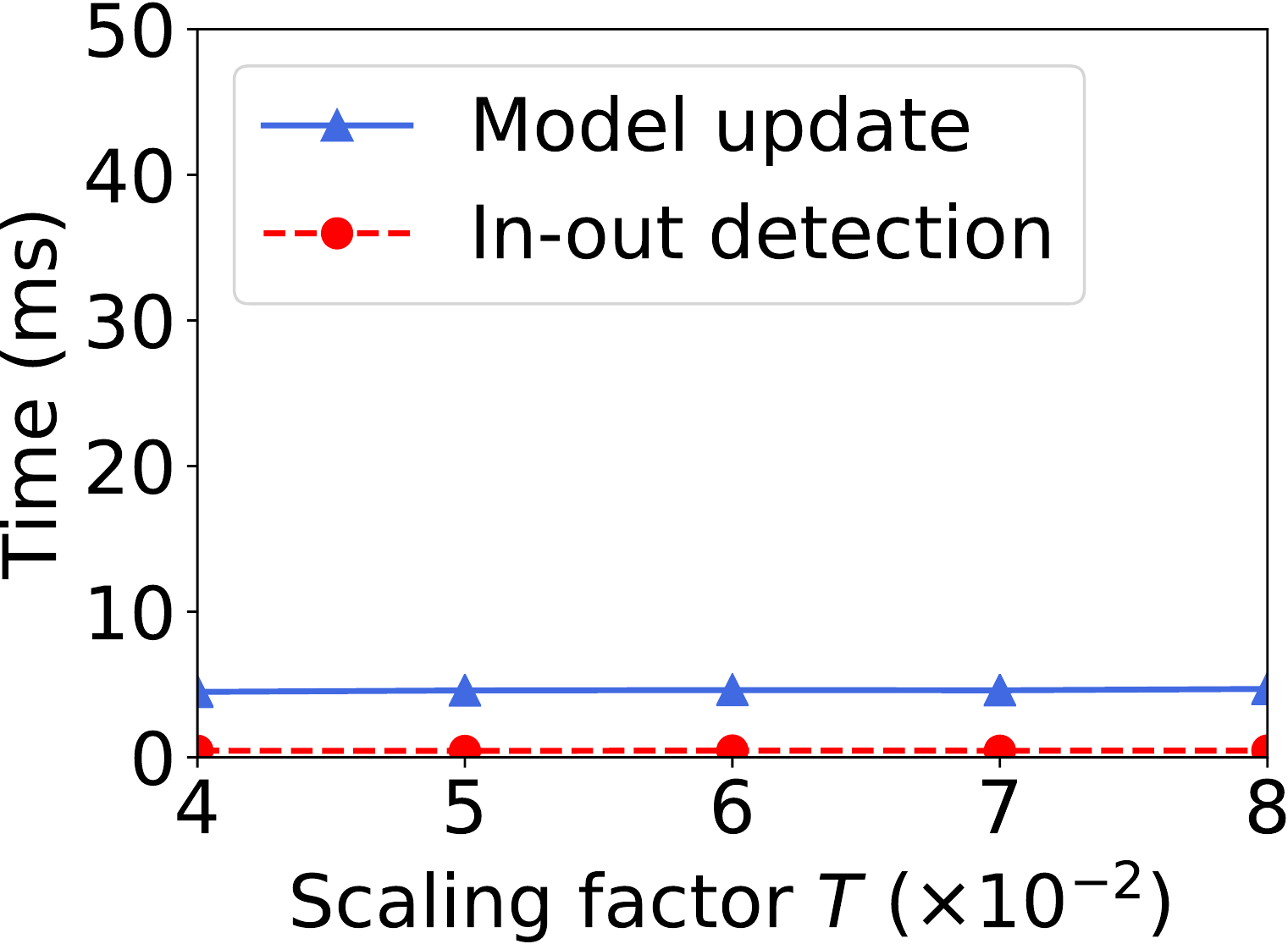}
    }
    \subfloat[]{%
    \includegraphics[width=0.2\linewidth]{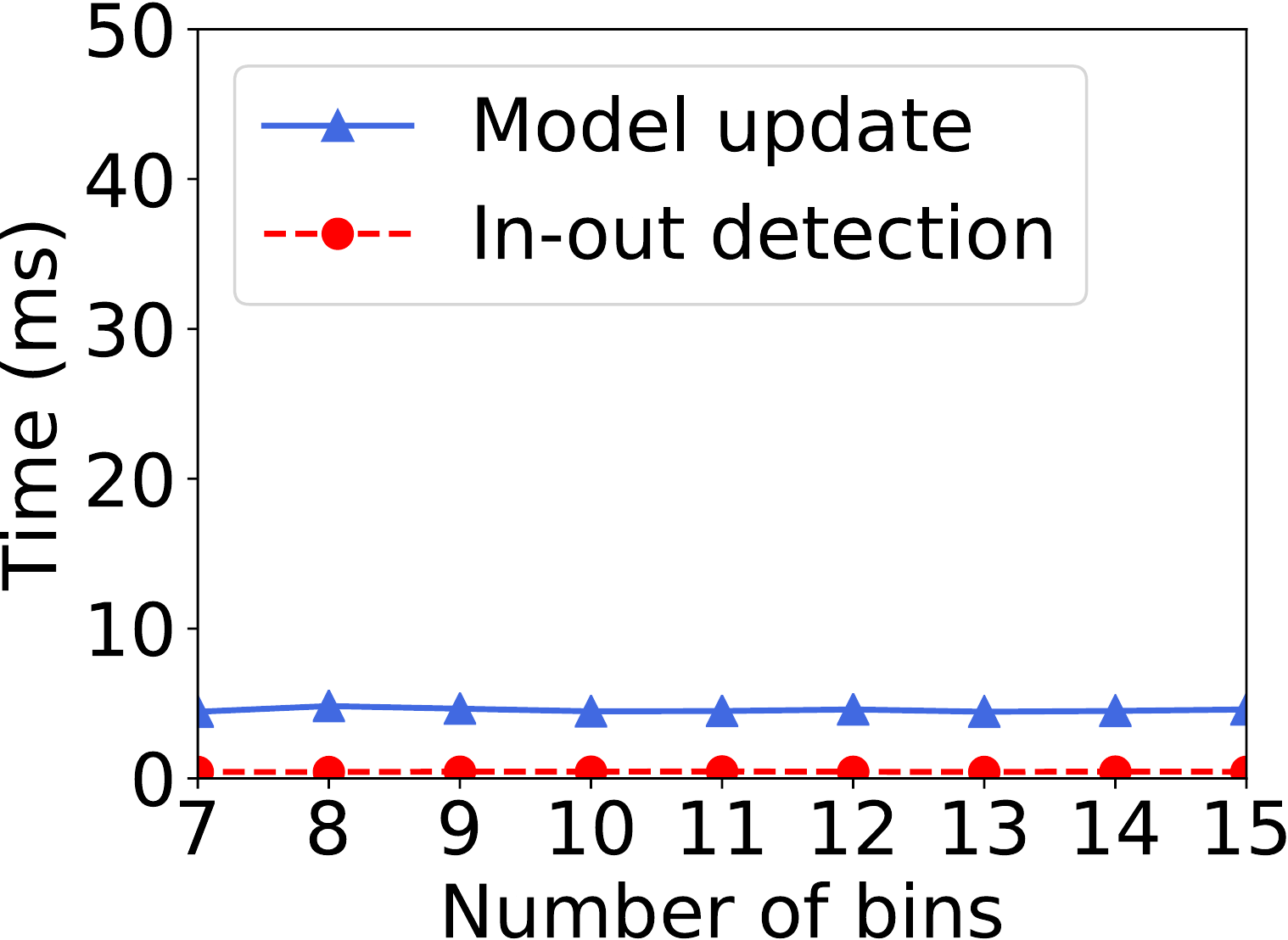}
    }
    \subfloat[]{%
    \includegraphics[width=0.2\linewidth]{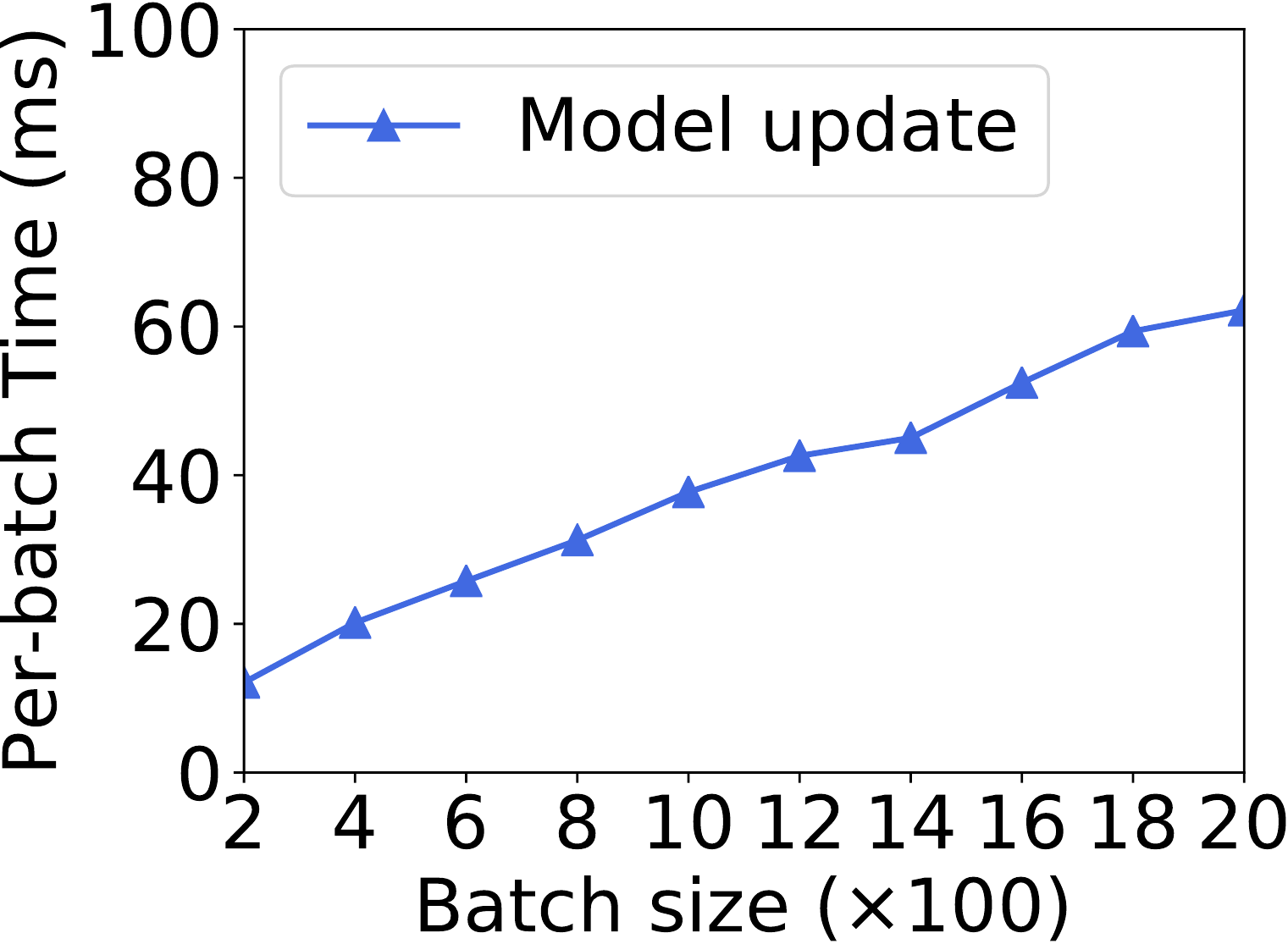}
    }
    \subfloat[]{%
    \includegraphics[width=0.19\linewidth]{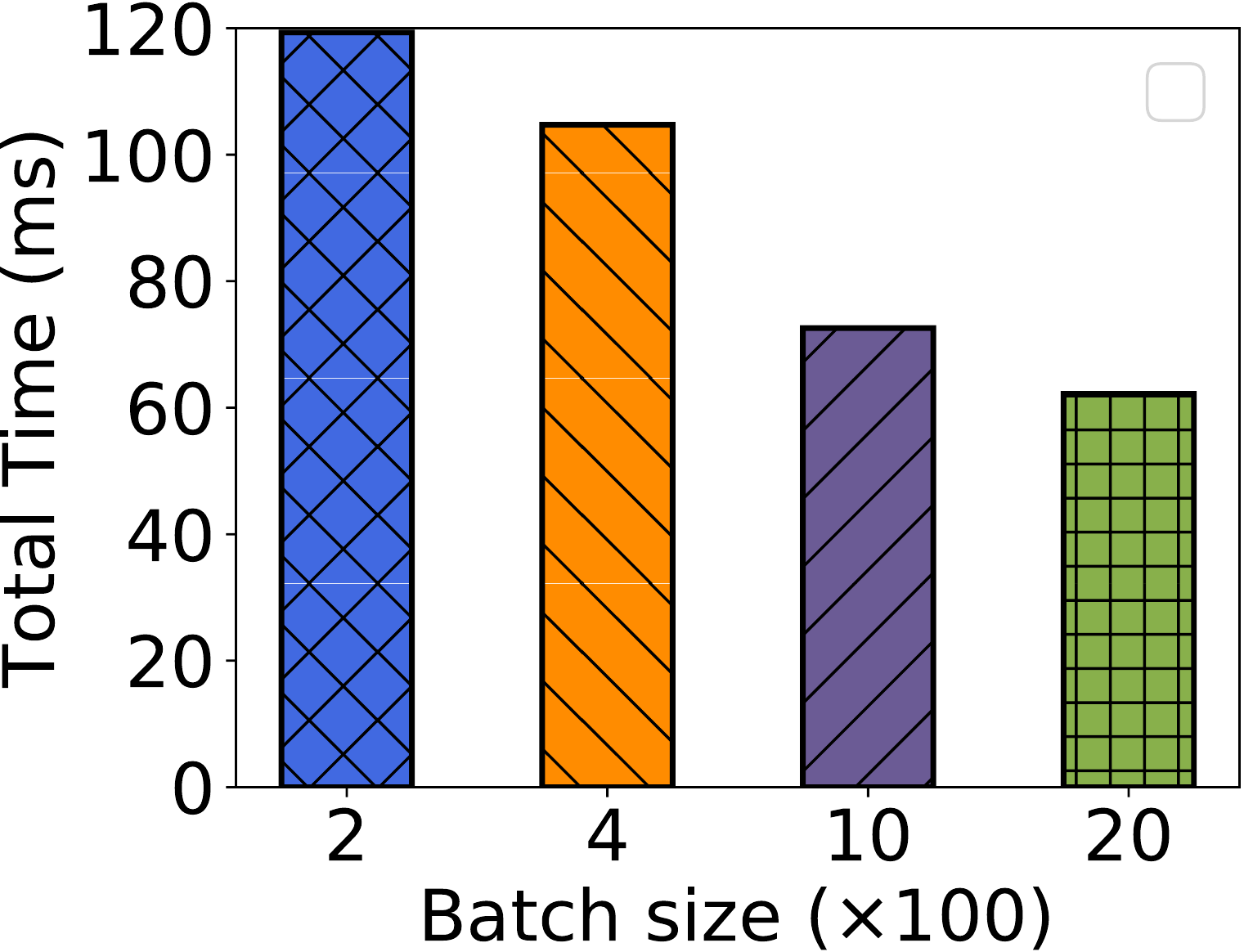}
    }
    \vspace{-0.05in}
	\caption{Running time versus (a) embedding dimension, (b) scaling factor $T$, and (c) bin size $m$; (d) Running time of processing a batch with different sizes; (e) Running time of processing 2000 samples with different batch sizes, e.g., 10 batches need to be processed for the batch size of 200.}
	\label{fig:running_time}
	\vspace{-0.1in}
\end{figure*}

\begin{figure*}[t]
    \centering
    \subfloat[Lab floorplan]{%
        \includegraphics[width=0.21\textwidth]{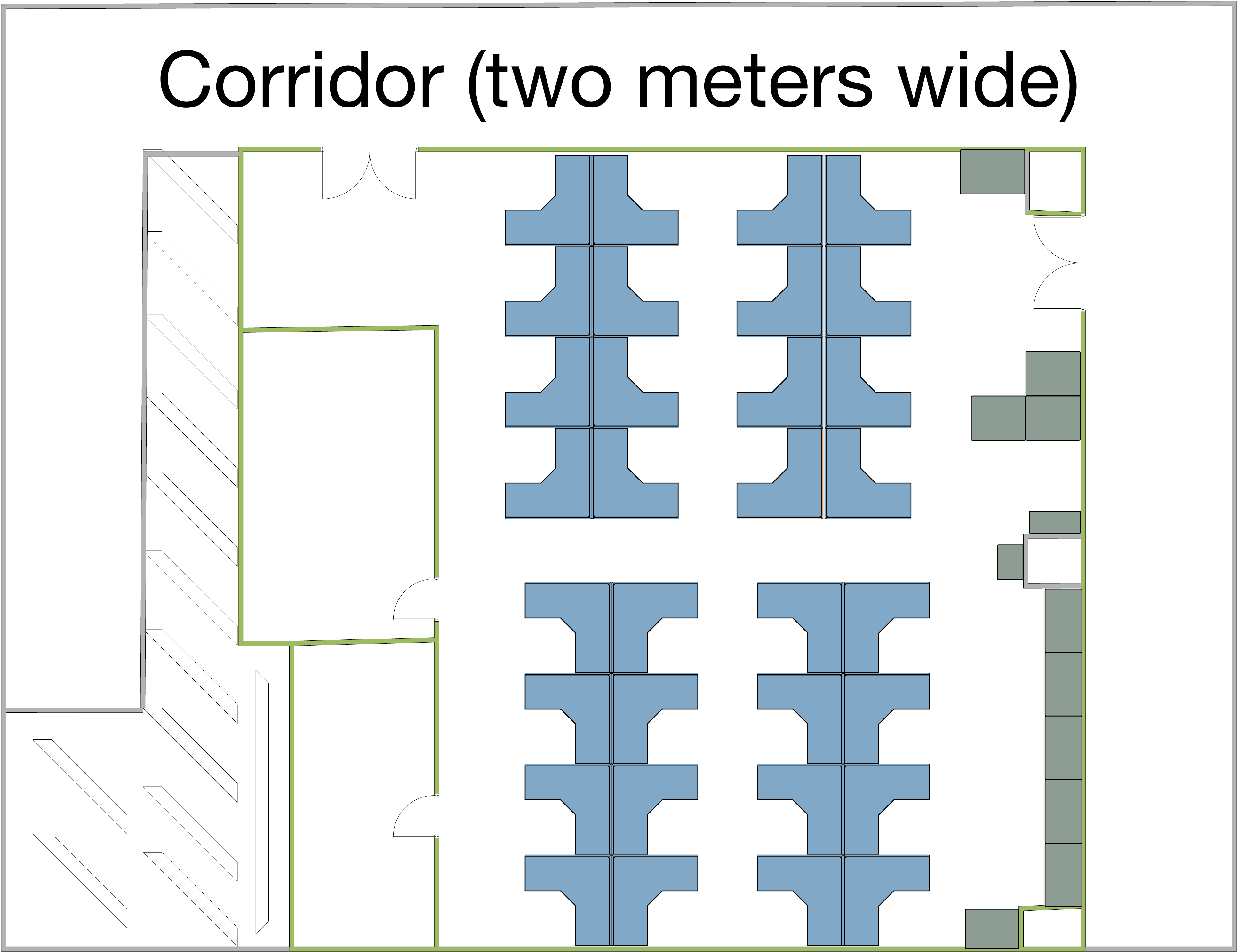}
        }
    \subfloat[Different time instances]{%
        \includegraphics[width=0.23\textwidth]{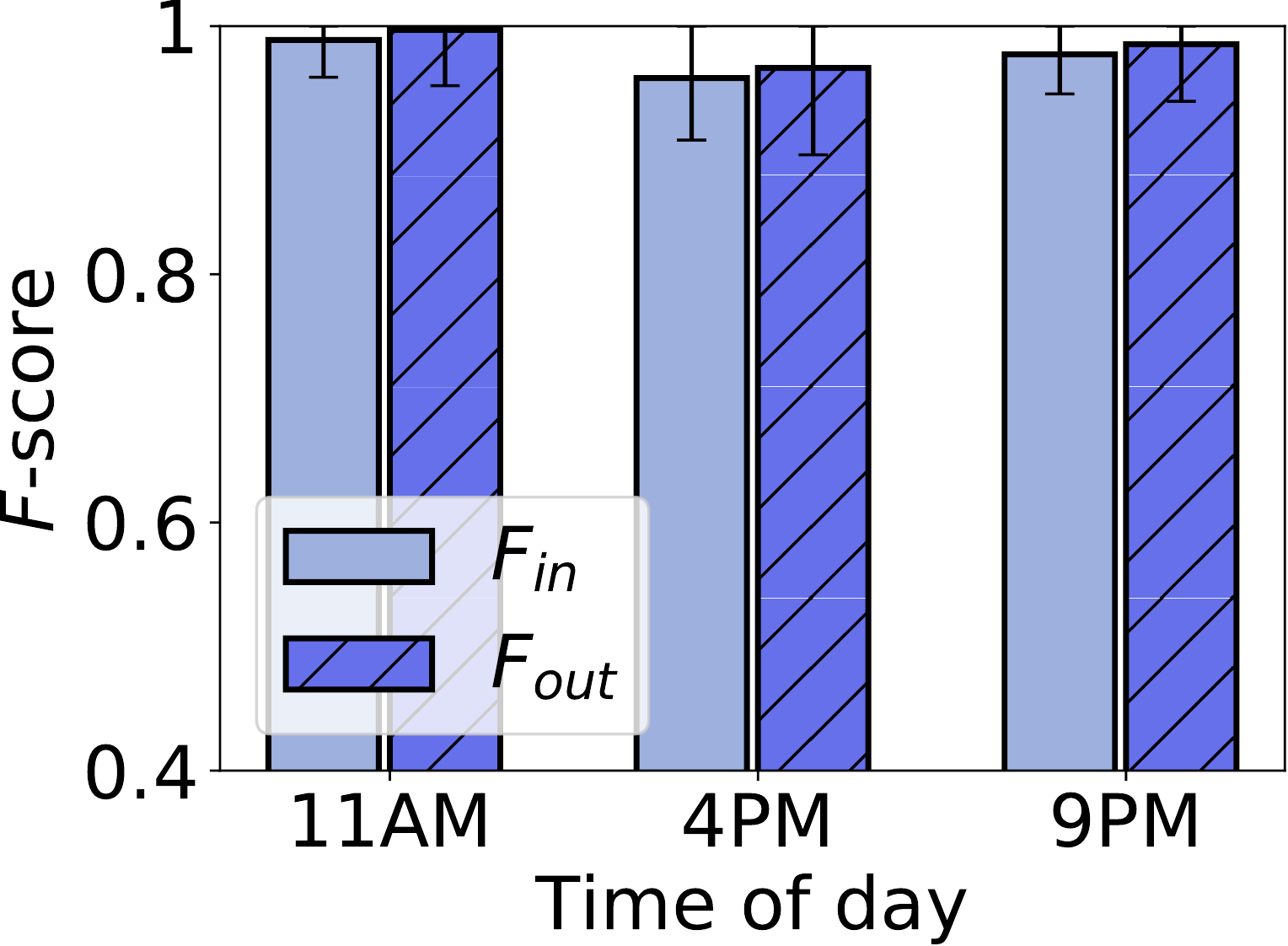}
        }
    \subfloat[Walking speed variations]{%
        \includegraphics[width=0.23\textwidth]{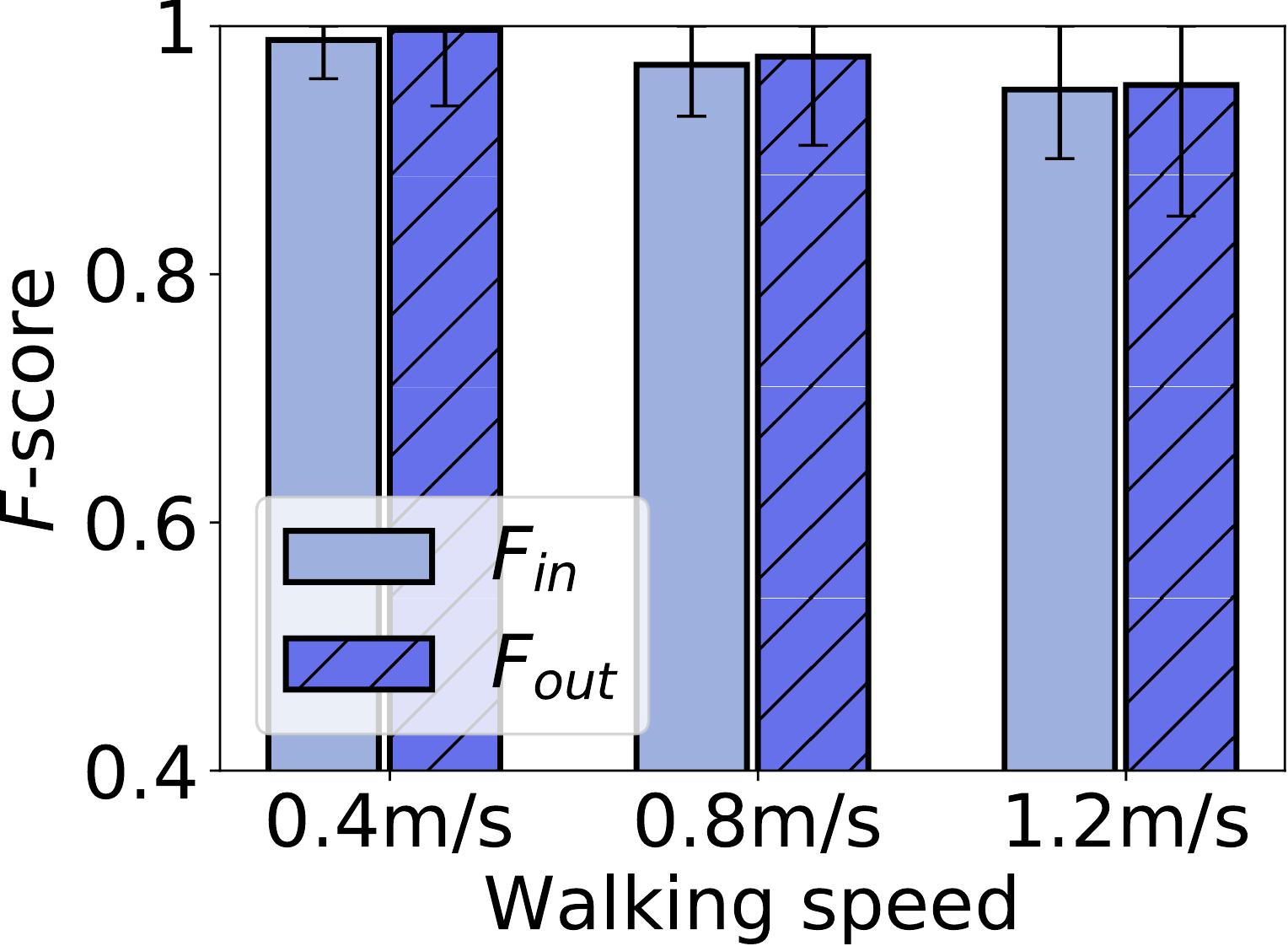}
        }
	\subfloat[Different frequency bands]{%
        \includegraphics[width=0.23\textwidth]{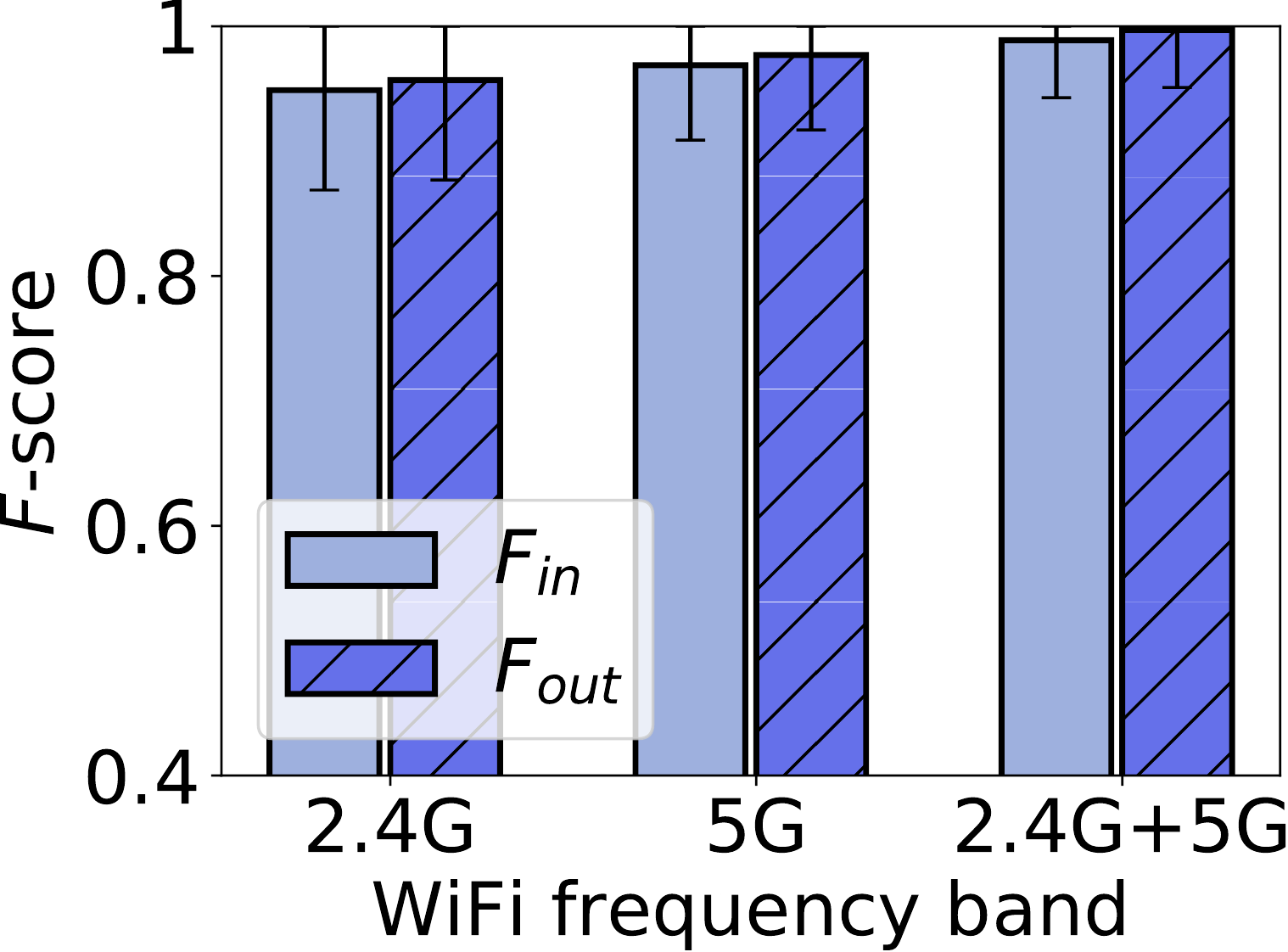}
        }
    \vspace{-0.05in}
	\caption{Lab experiments.}
	\label{fig:lab_exp}	
 	\vspace{-0.1in}
\end{figure*}

\vspace{2pt}
\noindent\textbf{Inference time measurement:} It is also important to see if \n{} is able to make the in-out decision \emph{in a prompt manner} upon arrival of a new RF signal record. To this end, we measure how long the entire inference process takes and obtain the breakdown of the inference time. Recall that the inference process has three steps, namely (1) obtaining the embeddings of the new RF record using \bisage{}, (2) in-out detection by our outlier detection algorithm, and (3) online model update with the new record. All results here are obtained by taking the average from 2000 runs.

We first show the impact of embedding dimension $d$ on the inference time in Figure~\ref{fig:running_time}(a). Since the embedding dimension affects all three steps in the inference process, we provide their running times. As shown in Figure~\ref{fig:running_time}(a), the embedding dimension has little impact on the running times of BiSAGE inference and in-out detection, while the running time of model update increases with $d$. It is because the model update involves updating $d$ histograms. A higher dimension indicates more histograms to update, thereby leading to a longer time for the update. Nonetheless, the total inference time is still less than 20 milliseconds even when $d \!=\! 128$. In addition, we evaluate the impact of scaling factor $T$ and the number of bins $m$ on the inference time. Since they only affect the inference process (i.e., in-out detection and model update) after the embeddings are learned by \bisage{}, we only report the running times of in-out detection and model update in Figure~\ref{fig:running_time}(b) and Figure~\ref{fig:running_time}(c). We see that $T$ and $m$ have little impact on the inference time. This again confirms the computationally inexpensive inference process of \n{}, which is suitable for real-time in-out detection.

We further discuss the impact of batch size on the inference time when we update our outlier detection model in batch mode. Recall that our model is updated only with \emph{highly confident} signal records. Since the batch update only affects the model update in the inference process, we only report its running time in Figure~\ref{fig:running_time}(d) and Figure~\ref{fig:running_time}(e). We first evaluate how long it takes to update the model with \emph{one} batch of (highly confident) signal records. As shown in Figure~\ref{fig:running_time}(d), the running time increases with the batch size since it takes longer time to recompute histograms by including a larger number of new records. We also evaluate the running time of updating the model using a total of 2000 signal records with different batch sizes. For example, we have ten batches with the batch size of 200, while having two batches with the batch size of 1000. As shown in Figure~\ref{fig:running_time}(e), the running time of model update decreases with increasing batch size as a reduction in the number of batches outweighs a saving in the model-update time with a smaller batch size. Thus, in practice, we can resort to the model update in batch mode.

\subsection{Impact of Environmental Factors}
\label{sec:exp_further}
We carry out further experiments to study the impact of environmental factors, such as time-varying RSS values, the variation in the walking speed of each user for collecting initial training data, and less availability of WiFi frequency bands, on the performance of \n{}. Each experiment is conducted once a day for three consecutive days in a lab environment (see Figure~\ref{fig:lab_exp}(a)). Note that the testing data for each experiment covers the area inside the lab (the geofencing area) and the corridor outside, which is only two meters wide, i.e. the boundary space of the geofencing area.

\vspace{2pt}
\noindent{\textbf{RSS variations over time.}} RF signal strength varies at different times of the day, mainly due to the people and their devices in the area. We select three representative time instances, i.e., 11AM, 4PM, and 9PM to test \n{}'s performance. At 11AM and 4PM, many people walk around the lab, and thus the signal strength may vary significantly. While at 9PM, signals would be more stable as the environments get quieter. Table~\ref{tab:rss_variation} summarizes the statistics of RSS values measured during a day, where SD stands for standard deviation. We collect the initial training data at 11AM and then conduct 50 walks around the lab for testing \n{} at each time instance. Thanks to the model updating mechanism, \n{} can adapt to time-varying RSS values and thus achieve outstanding performance consistently, as shown in Figure~\ref{fig:lab_exp}(b). In other words, \n{} remains effective even in such a dynamic RF signal environment.
\begin{table}[h]
\vspace{-0.1in}
\caption{RSS variation during a day.}
\vspace{-0.05in}
\centering 
\small
\begin{tabular}{|c | c | c | c |}
\hline
Time & Mean (dBm) & SD (dBm) & \#MACs \\
\hline
11~AM & $-67.76$ & 8.63 & 68\\
\hline
4~PM & $-81.67$ & 13.50 & 85\\
\hline
9~PM & $-78.64$ & 11.03 & 47\\
\hline
\end{tabular}
\vspace{-0.1in}
\label{tab:rss_variation}
\end{table}

\vspace{2pt}
\noindent\textbf{Walking speed variations:} For initial training, each user is asked to walk around the house for a short period of time. By varying the user's walking speed, we carry out three different experiments, i.e., slow walk ($\sim\! 0.4\text{m/s}$), normal walk ($\sim\! 0.8\text{m/s}$), and fast walk ($\sim\! 1.2\text{m/s}$). For each experiment, we walk along the inner perimeter of the lab twice for initial training. For testing purposes, we conduct 50 walks around (inside and outside) the lab. As shown in Figure~\ref{fig:lab_exp}(c), \n{} exhibits excellent performance in all three cases.

\vspace{2pt}
\noindent\textbf{Different frequency-band availabilities:} We study the geo-fencing performance versus APs' frequency bands as more devices support dual and tri-bands. Here, we train and test the model with signals of 2.4GHz only, 5GHz only, and 2.4GHz~+~5GHz. As presented in Figure~\ref{fig:lab_exp}(d), more frequency bands (i.e., 2.4GHz~+~5GHz) improve the performance, even though our system remains effective in the 2.4GHz only environment. We also observe that a higher frequency band results in better performance, as higher frequency signals are more likely confined within the lab space.

\subsection{Model Scalability}

\begin{figure}[h]
    \centering
    \includegraphics[width=0.48\textwidth]{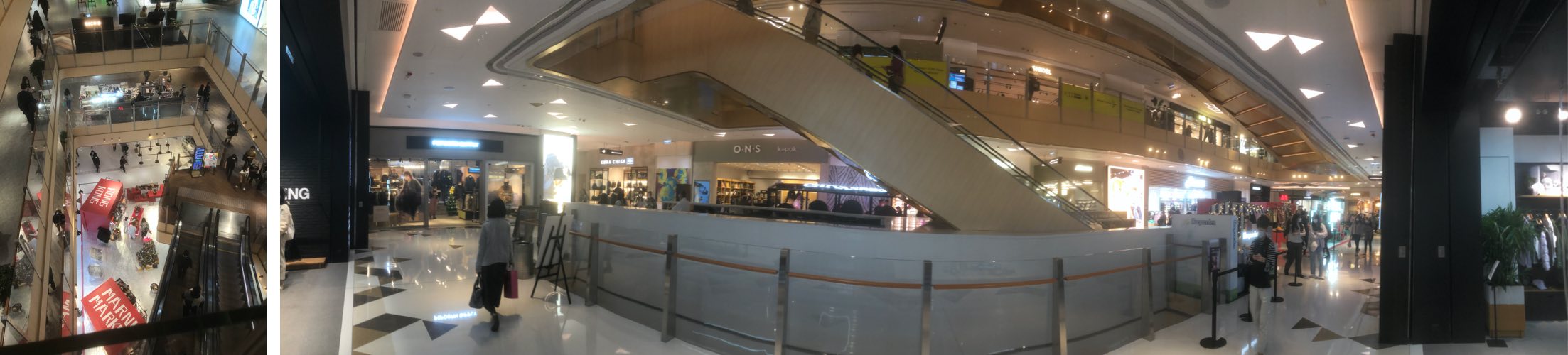}
    \caption{Shopping mall layout.}
    \label{fig:mall}
\end{figure}

To validate the scalability of our model with large-scale datasets, we conduct additional experiments in a five-story shopping mall (Figure~\ref{fig:mall}) and based on the UJI open dataset~\cite{torres2014ujiindoorloc} from Kaggle competition, which covers three buildings and contains 21,048 RF signal records. For each building, we define the middle floor as the geofenced area and the other floors as outside. In the shopping-mall experiment, we walk around the third floor (middle floor) to collect about 5,000 signal records for initial training. We then walk randomly within the five-story building to collect about 200,000 signal records for testing. For the experiment on the UJI dataset (per building), we uniformly sample half of the signal records from the middle floor (i.e., around 800 records) for initial training and use the rest of the records for testing. The testing data are streamed in the same way as in the former experiments, meaning that the datasets are treated as dynamic ones. In other words, our model is again being updated during the testing process.
\begin{table}[t]
\caption{Performance comparison in the shopping mall dataset.}
\vspace{-0.1in}
\centering 
\resizebox{0.5\textwidth}{!}{
\begin{tabular}{l c c c c c c c c c c}
\toprule
 \multirow{2}{*}{Algorithms}&\hspace{0.1in}& \multicolumn{3}{c}{In-premises detection} &\hspace{0.1in}& \multicolumn{3}{c}{Outside detection} \\
 && $P_{in}$ & $R_{in}$ & $F_{in}$ & \hspace{0.1in} & $P_{out}$ & $R_{out}$ & $F_{out}$ \\
\toprule
\n{} && \textbf{0.94}  & \textbf{0.99} & \textbf{0.96} &\hspace{0.1in}& \textbf{0.98} & \textbf{0.96} & \textbf{0.97} \\
\hline
SignatureHome && 0.68 & 0.85  & 0.75 &\hspace{0.1in}& 0.76 & 0.72 & 0.74\\
\hline
INOA && 0.75 & 0.89 & 0.81 &\hspace{0.1in}& 0.80 & 0.78 & 0.79 \\
\bottomrule
\end{tabular}
}
\label{tab:results_mall}
\vspace{-0.1in}
\end{table}

\begin{table}[t]
\caption{Performance comparison in Building 0 of the UJI open dataset.}
\vspace{-0.1in}
\centering 
\resizebox{0.5\textwidth}{!}{
\begin{tabular}{l c c c c c c c c c c}
\toprule
 \multirow{2}{*}{Algorithms}&\hspace{0.1in}& \multicolumn{3}{c}{In-premises detection} &\hspace{0.1in}& \multicolumn{3}{c}{Outside detection} \\
 && $P_{in}$ & $R_{in}$ & $F_{in}$ & \hspace{0.1in} & $P_{out}$ & $R_{out}$ & $F_{out}$ \\
\toprule
\n{} && \textbf{0.91}  & \textbf{0.99} & \textbf{0.95} &\hspace{0.1in}& \textbf{0.99} & \textbf{0.98} & \textbf{0.99} \\
\hline
SignatureHome && 0.62 & 0.87  & 0.72 &\hspace{0.1in}& 0.85 & 0.83 & 0.84 \\
\hline
INOA && 0.68 & 0.89 & 0.77 &\hspace{0.1in}& 0.88 & 0.85 & 0.86 \\
\bottomrule
\end{tabular}
}
\label{tab:results_uji0}
\vspace{-0.1in}
\end{table}

We compare \n{} with SignatureHome and INOA and report the results in Tables~\ref{tab:results_mall}--\ref{tab:results_uji2}. \n{} outperforms SignatureHome and INOA significantly. It is because the embeddings learned by \bisage{} well preserve the relative proximity between signal records, and our enhanced outlier detection algorithm accurately predicts outliers from normal data samples. However, SignatureHome does not benefit from an associated IP within the geofencing area as the experiments are done based only on RF signals. Moreover, since signals from an AP/MAC can be detected on multiple floors, a mere calculation of the overlap ratio of MACs in SignatureHome can hardly differentiate the signal records collected within the geofencing area form the ones outside. In addition, although INOA performs better than SignatureHome, it also suffers from the same problem. The nature of RF signals from an AP/MAC that can reach different floors makes the learning from the signal records for each pair of APs/MACs insufficient for accurate in-out detection. 

\begin{table}[t]
\caption{Performance comparison in Building 1 of the UJI open dataset.}
\vspace{-0.1in}
\centering 
\resizebox{0.5\textwidth}{!}{
\begin{tabular}{l c c c c c c c c c c}
\toprule
 \multirow{2}{*}{Algorithms}&\hspace{0.1in}& \multicolumn{3}{c}{In-premises detection} &\hspace{0.1in}& \multicolumn{3}{c}{Outside detection} \\
 && $P_{in}$ & $R_{in}$ & $F_{in}$ & \hspace{0.1in} & $P_{out}$ & $R_{out}$ & $F_{out}$ \\
\toprule
\n{} && \textbf{0.86}  & \textbf{0.99} & \textbf{0.92} &\hspace{0.1in}& \textbf{0.99} & \textbf{0.97} & \textbf{0.98} \\
\hline
SignatureHome && 0.56 & 0.84  & 0.67 &\hspace{0.1in}& 0.79 & 0.78 & 0.78 \\
\hline
INOA && 0.65 & 0.88 & 0.75 &\hspace{0.1in}& 0.86 & 0.82 & 0.84 \\
\bottomrule
\end{tabular}
}
\label{tab:results_uji1}
\vspace{-0.1in}
\end{table}

\begin{table}[t]
\caption{Performance comparison in Building 2 of the UJI open dataset.}
\vspace{-0.1in}
\centering 
\resizebox{0.5\textwidth}{!}{
\begin{tabular}{l c c c c c c c c c c}
\toprule
 \multirow{2}{*}{Algorithms}&\hspace{0.1in}& \multicolumn{3}{c}{In-premises detection} &\hspace{0.1in}& \multicolumn{3}{c}{Outside detection} \\
 && $P_{in}$ & $R_{in}$ & $F_{in}$ & \hspace{0.1in} & $P_{out}$ & $R_{out}$ & $F_{out}$ \\
\toprule
\n{} && \textbf{0.84}  & \textbf{0.99} & \textbf{0.91} &\hspace{0.1in}& \textbf{0.99} & \textbf{0.98} & \textbf{0.99} \\
\hline
SignatureHome && 0.50 & 0.82  & 0.62 &\hspace{0.1in}& 0.82 & 0.77 & 0.79 \\
\hline
INOA && 0.59 & 0.85 & 0.69 &\hspace{0.1in}& 0.85 & 0.82 & 0.83 \\
\bottomrule
\end{tabular}
}
\label{tab:results_uji2}
\vspace{-0.1in}
\end{table}

\section{Discussion}
\label{sec:discuss}
We discuss in this section the application scenarios of \n{} and further elaborate on how it is different from fingerprinting-based indoor localization techniques. We also discuss possible attacks on the system.

Our data are collected using wireless devices with moderate RF sensing frequencies ($\sim \!1$Hz). In other words, a user is expected to move for one or two meters between two consecutive sensing events. Hence, our proposed system \n{} is intended for geofencing applications requiring precision at the \emph{meter} level, such as elderly care and restricted UAV navigation. The goal is to alert when the user gets out of the geofenced area, which is enabled by our effective representation learning on RF signals. Nonetheless, geofencing applications requiring higher granularity would always be interesting. We consider the more strict geofencing problems at the \emph{sub-meter} level as a future research direction.

One may consider fingerprinting-based indoor localization techniques as a feasible solution to our geofencing problem. They generally require the user to build fingerprints, i.e., RF signals at predetermined specific locations, to provide an estimate of the user’s location within the area of interest. However, their application to in-out detection is not straightforward. Since they merely provide an estimate of the user’s location, the \emph{map} of the geofencing area should be available for the location estimate to be usable for in-out detection. The fingerprints also need to be collected at carefully chosen locations both \emph{inside} and \emph{outside} the geofencing area. In contrast, \n{} does \emph{not} require any indoor maps or floor plans but only needs RF signals collected \emph{inside} the geofencing area.

In addition, one may be concerned about the vulnerability of our online model update scheme to some potential attack by `bad actors' who intentionally stay at the boundary and move outward slowly to abuse the online model update and thus break the system. However, this is \emph{hardly} possible in practice. In a typical geofencing scenario such as nursing home or medical observation, the boundary of the premises is usually well defined and clear, i.e., the geofenced region and outside are often separated by walls or partitions. Such physical objects bounce RF signals off and attenuate the signals, e.g., a signal attenuation of 3 dB for drywalls (half signal strength) and up to 10 dB for brick walls (10 times less signal strength), thereby making the signal characteristics inside and outside quite different. In other words, the physical objects not only obstruct the bad actors in staying at the boundary but also prevent them from building up a collection of RF signals whose strengths are gradually changing. In addition, \n{} only leverages highly confident samples (around 5\% of newly predicted in-premises samples) to update the `signal boundary'. The signals collected outside the boundary are quite unlikely to be used for our online model update.
\vspace{0.08in}

\section{Conclusion}
\label{sec:conclude}
We have proposed \n{}, a practical yet effective geofencing system that only leverages ambient RF signals without any extra hardware or continuous intervention from users. \n{} is built upon three integral components, i.e., a representation of RF signals via a weighted bipartite graph, our novel bipartite network embedding algorithm \bisage{}, and our enhanced histogram-based detection algorithm. They enable \n{} to achieve highly accurate in-out detection performance while being robust to dynamic RF environments. We have empirically demonstrated the robustness and superior performance of \n{} in various housing and RF scenarios.

We expect that the key enablers of \n{} have broader impacts. The bipartite graph modeling and \bisage{} could be used as a viable solution to dealing with \emph{variable-length} feature vectors for building a learning model. \bisage{} could also be applied for network embedding and representation learning on \emph{general} bipartite graphs with other applications.

\bibliography{ref}

\begin{thebibliography}{10}
\providecommand{\url}[1]{#1}
\csname url@samestyle\endcsname
\providecommand{\newblock}{\relax}
\providecommand{\bibinfo}[2]{#2}
\providecommand{\BIBentrySTDinterwordspacing}{\spaceskip=0pt\relax}
\providecommand{\BIBentryALTinterwordstretchfactor}{4}
\providecommand{\BIBentryALTinterwordspacing}{\spaceskip=\fontdimen2\font plus
\BIBentryALTinterwordstretchfactor\fontdimen3\font minus
  \fontdimen4\font\relax}
\providecommand{\BIBforeignlanguage}[2]{{%
\expandafter\ifx\csname l@#1\endcsname\relax
\typeout{** WARNING: IEEEtran.bst: No hyphenation pattern has been}%
\typeout{** loaded for the language `#1'. Using the pattern for}%
\typeout{** the default language instead.}%
\else
\language=\csname l@#1\endcsname
\fi
#2}}
\providecommand{\BIBdecl}{\relax}
\BIBdecl

\bibitem{helmy2016alzimio}
J.~Helmy and A.~Helmy, ``The alzimio app for dementia, autism amp; alzheimer's:
  Using novel activity recognition algorithms and geofencing,'' in \emph{IEEE
  SMARTCOMP}, 2016.

\bibitem{SignatureHome}
J.~Tan, E.~Sumpena, W.~Zhuo, Z.~Zhao, M.~Liu, and S.-H.~G. Chan, ``Iot
  geofencing for covid-19 home quarantine enforcement,'' \emph{IEEE Internet of
  Things Magazine}, vol.~3, no.~3, pp. 24--29, 2020.

\bibitem{hermand2018constrained}
E.~Hermand, T.~W. Nguyen, M.~Hosseinzadeh, and E.~Garone, ``Constrained control
  of uavs in geofencing applications,'' in \emph{IEEE MED}, 2018.

\bibitem{oliveira2015intelligent}
R.~R. Oliveira, I.~M. Cardoso, J.~L. Barbosa, C.~A. da~Costa, and M.~P. Prado,
  ``An intelligent model for logistics management based on geofencing
  algorithms and rfid technology,'' \emph{Expert Syst. Appl.}, 2015.

\bibitem{greenwald2011economically}
A.~Greenwald, G.~Hampel, C.~Phadke, and V.~Poosala, ``An economically viable
  solution to geofencing for mass-market applications,'' \emph{Bell Labs Tech.
  J.}, 2011.

\bibitem{zang2010bayesian}
H.~Zang, F.~Baccelli, and J.~Bolot, ``Bayesian inference for localization in
  cellular networks,'' in \emph{IEEE INFOCOM}, 2010.

\bibitem{nirjon2014coin}
S.~Nirjon, J.~Liu, G.~DeJean, B.~Priyantha, Y.~Jin, and T.~Hart, ``Coin-gps:
  Indoor localization from direct gps receiving,'' in \emph{MobiSys}, 2014.

\bibitem{schloemann2015toward}
J.~Schloemann, H.~S. Dhillon, and R.~M. Buehrer, ``Toward a tractable analysis
  of localization fundamentals in cellular networks,'' \emph{IEEE Trans. Wirel.
  Commun.}, 2015.

\bibitem{rizk2018cellindeep}
H.~Rizk, M.~Torki, and M.~Youssef, ``Cellindeep: Robust and accurate
  cellular-based indoor localization via deep learning,'' \emph{IEEE Sens. J.},
  2018.

\bibitem{margolies2017can}
R.~Margolies, R.~Becker, S.~Byers, S.~Deb, R.~Jana, S.~Urbanek, and
  C.~Volinsky, ``Can you find me now? evaluation of network-based localization
  in a 4g lte network,'' in \emph{IEEE INFOCOM}, 2017.

\bibitem{hoang2019recurrent}
M.~T. Hoang, B.~Yuen, X.~Dong, T.~Lu, R.~Westendorp, and K.~Reddy, ``Recurrent
  neural networks for accurate rssi indoor localization,'' \emph{IEEE Internet
  Things J.}, vol.~6, no.~6, pp. 10\,639--10\,651, 2019.

\bibitem{li2019smartloc}
L.~Li, X.~Guo, and N.~Ansari, ``Smartloc: Smart wireless indoor localization
  empowered by machine learning,'' \emph{IEEE Trans. Ind. Electron.}, 2019.

\bibitem{wang2020spatial}
R.~Wang, H.~Luo, Q.~Wang, Z.~Li, F.~Zhao, and J.~Huang, ``A spatial--temporal
  positioning algorithm using residual network and lstm,'' \emph{IEEE Trans
  Instrum Meas}, 2020.

\bibitem{zhou2021integrated}
M.~Zhou, Y.~Li, M.~J. Tahir, X.~Geng, Y.~Wang, and W.~He, ``Integrated
  statistical test of signal distributions and access point contributions for
  wi-fi indoor localization,'' \emph{IEEE Trans. Veh. Technol.}, 2021.

\bibitem{fan2021siabr}
S.~Fan, Y.~Wu, C.~Han, and X.~Wang, ``Siabr: A structured intra-attention
  bidirectional recurrent deep learning method for ultra-accurate terahertz
  indoor localization,'' \emph{IEEE J. Sel. Areas Commun.}, 2021.

\bibitem{chen2022fidora}
X.~Chen, H.~Li, C.~Zhou, X.~Liu, D.~Wu, and G.~Dudek, ``Fidora: Robust
  wifi-based indoor localization via unsupervised domain adaptation,''
  \emph{IEEE Internet Things J.}, 2022.

\bibitem{HBOS}
M.~Goldstein and A.~Dengel, ``Histogram-based outlier score (hbos): A fast
  unsupervised anomaly detection algorithm,'' in \emph{KI-2012}, 2012.

\bibitem{taxSupportVectorData2004}
D.~M.~J. Tax and R.~P.~W. Duin, ``Support vector data description,''
  \emph{Mach. Learn.}, 2004.

\bibitem{chalapathy2018anomaly}
R.~Chalapathy, A.~K. Menon, and S.~Chawla, ``Anomaly detection using one-class
  neural networks,'' \emph{arXiv preprint}, 2018.

\bibitem{DeepSVDD18}
L.~Ruff, R.~Vandermeulen, N.~Goernitz, L.~Deecke, S.~A. Siddiqui, A.~Binder,
  E.~M{\"u}ller, and M.~Kloft, ``Deep one-class classification,'' in
  \emph{ICML}, 2018.

\bibitem{goyal2020drocc}
S.~Goyal, A.~Raghunathan, M.~Jain, H.~V. Simhadri, and P.~Jain, ``Drocc: Deep
  robust one-class classification,'' in \emph{ICML}, 2020.

\bibitem{zaheer2020old}
M.~Z. Zaheer, J.-H. Lee, M.~Astrid, and S.-I. Lee, ``Old is gold: Redefining
  the adversarially learned one-class classifier training paradigm,'' in
  \emph{IEEE CVPR}, 2020.

\bibitem{liznerski2021explainable}
P.~Liznerski, L.~Ruff, R.~A. Vandermeulen, B.~J. Franks, M.~Kloft, and K.-R.
  M{\"u}ller, ``Explainable deep one-class classification,'' in \emph{ICLR},
  2021.

\bibitem{graphsage}
W.~Hamilton, Z.~Ying, and J.~Leskovec, ``Inductive representation learning on
  large graphs,'' in \emph{NeurIPS}, 2017.

\bibitem{INOA18}
K.-H. Chow, S.~He, J.~Tan, and S.-H.~G. Chan, ``Efficient locality
  classification for indoor fingerprint-based systems,'' \emph{IEEE Trans.
  Mobile Comput.}, 2019.

\bibitem{boysen2014constructing}
M.~Boysen, C.~de~Haas, H.~Lu, X.~Xie, and A.~Pilvinyte, ``Constructing indoor
  navigation systems from digital building information,'' in \emph{2014 IEEE
  30th International Conference on Data Engineering (ICDE)}.\hskip 1em plus
  0.5em minus 0.4em\relax IEEE, 2014.

\bibitem{li2016vita}
H.~Li, H.~Lu, X.~Chen, G.~Chen, K.~Chen, and L.~Shou, ``Vita: A versatile
  toolkit for generating indoor mobility data for real-world buildings,''
  \emph{Proceedings of the VLDB Endowment}, vol.~9, no.~13, pp. 1453--1456,
  2016.

\bibitem{baba2016learning}
A.~I. Baba, M.~Jaeger, H.~Lu, T.~B. Pedersen, W.-S. Ku, and X.~Xie,
  ``{Learning-based cleansing for indoor RFID data},'' in \emph{ACM SIGMOD},
  2016.

\bibitem{lin2021locater}
Y.~Lin and D.~Jiang, ``{LOCATER}: Cleaning {WiFi} connectivity datasets for
  semantic localization,'' \emph{Proceedings of the VLDB Endowment}, 2021.

\bibitem{gao2018bine}
M.~Gao, L.~Chen, X.~He, and A.~Zhou, ``Bine: Bipartite network embedding,'' in
  \emph{ACM SIGIR}, 2018.

\bibitem{GRAFICS}
W.~Zhuo, Z.~Zhao, K.~H. Chiu, S.~Li, S.~Ha, C.-H. Lee, and S.-H.~G. Chan,
  ``{GRAFICS}: {Graph} {Embedding}-based {Floor} {Identification} {Using}
  {Crowdsourced} {RF} {Signals},'' in \emph{ICDCS}, 2022.

\bibitem{you2020handling}
J.~You, X.~Ma, Y.~Ding, M.~J. Kochenderfer, and J.~Leskovec, ``Handling missing
  data with graph representation learning,'' in \emph{NeurIPS}, 2020.

\bibitem{dong2017metapath2vec}
Y.~Dong, N.~V. Chawla, and A.~Swami, ``metapath2vec: Scalable representation
  learning for heterogeneous networks,'' in \emph{ACM KDD}, 2017.

\bibitem{shi2018heterogeneous}
C.~Shi, B.~Hu, W.~X. Zhao, and S.~Y. Philip, ``Heterogeneous information
  network embedding for recommendation,'' \emph{IEEE Transactions on Knowledge
  and Data Engineering}, vol.~31, no.~2, pp. 357--370, 2018.

\bibitem{fu2020magnn}
X.~Fu, J.~Zhang, Z.~Meng, and I.~King, ``Magnn: Metapath aggregated graph
  neural network for heterogeneous graph embedding,'' in \emph{WWW}, 2020.

\bibitem{chatzopoulos2020sphinx}
S.~Chatzopoulos, K.~Patroumpas, A.~Zeakis, T.~Vergoulis, and D.~Skoutas,
  ``Sphinx: A system for metapath-based entity exploration in heterogeneous
  information networks,'' \emph{Proceedings of the VLDB Endowment}, vol.~13,
  no.~12, pp. 2913--2916, 2020.

\bibitem{fang2020effective}
Y.~Fang, Y.~Yang, W.~Zhang, X.~Lin, and X.~Cao, ``Effective and efficient
  community search over large heterogeneous information networks,''
  \emph{Proceedings of the VLDB Endowment}, vol.~13, no.~6, pp. 854--867, 2020.

\bibitem{wang2020dynamic}
X.~Wang, Y.~Lu, C.~Shi, R.~Wang, P.~Cui, and S.~Mou, ``Dynamic heterogeneous
  information network embedding with meta-path based proximity,'' \emph{IEEE
  Transactions on Knowledge and Data Engineering}, vol.~34, no.~3, pp.
  1117--1132, 2022.

\bibitem{yang2021interpretable}
Y.~Yang, Z.~Guan, J.~Li, W.~Zhao, J.~Cui, and Q.~Wang, ``Interpretable and
  efficient heterogeneous graph convolutional network,'' \emph{IEEE
  Transactions on Knowledge and Data Engineering}, 2021.

\bibitem{gu2022hybridgnn}
T.~Gu, C.~Wang, C.~Wu, Y.~Lou, J.~Xu, C.~Wang, K.~Xu, C.~Ye, and Y.~Song,
  ``{HybridGNN}: Learning hybrid representation for recommendation in multiplex
  heterogeneous networks,'' in \emph{2022 IEEE 38th International Conference on
  Data Engineering (ICDE)}.\hskip 1em plus 0.5em minus 0.4em\relax IEEE, 2022.

\bibitem{wang2019heterogeneous}
X.~Wang, H.~Ji, C.~Shi, B.~Wang, Y.~Ye, P.~Cui, and P.~S. Yu, ``Heterogeneous
  graph attention network,'' in \emph{WWW}, 2019.

\bibitem{yang2020multisage}
C.~Yang, A.~Pal, A.~Zhai, N.~Pancha, J.~Han, C.~Rosenberg, and J.~Leskovec,
  ``Multisage: Empowering gcn with contextualized multi-embeddings on web-scale
  multipartite networks,'' in \emph{ACM KDD}, 2020.

\bibitem{huang2022estimating}
C.~Huang, Y.~Fang, X.~Lin, X.~Cao, W.~Zhang, and M.~Orlowska, ``Estimating node
  importance values in heterogeneous information networks,'' in \emph{2022 IEEE
  38th International Conference on Data Engineering (ICDE)}.\hskip 1em plus
  0.5em minus 0.4em\relax IEEE, 2022.

\bibitem{kdd17Anomaly}
C.~Zhou and R.~C. Paffenroth, ``Anomaly detection with robust deep
  autoencoders,'' in \emph{ACM KDD}, 2017.

\bibitem{corain2021dbscout}
M.~Corain, P.~Garza, and A.~Asudeh, ``{DBSCOUT}: A density-based method for
  scalable outlier detection in very large datasets,'' in \emph{2021 IEEE 37th
  International Conference on Data Engineering (ICDE)}.\hskip 1em plus 0.5em
  minus 0.4em\relax IEEE, 2021.

\bibitem{kieu2022robust}
T.~Kieu, B.~Yang, C.~Guo, C.~S. Jensen, Y.~Zhao, F.~Huang, and K.~Zheng,
  ``Robust and explainable autoencoders for time series outlier detection,'' in
  \emph{2022 IEEE 38th International Conference on Data Engineering
  (ICDE)}.\hskip 1em plus 0.5em minus 0.4em\relax IEEE, 2022.

\bibitem{abbas2019wideep}
M.~Abbas, M.~Elhamshary, H.~Rizk, M.~Torki, and M.~Youssef, ``{{WiDeep}}:
  {{WiFi}}-based accurate and robust indoor localization system using deep
  learning,'' in \emph{{{IEEE PerCom}}}, 2019.

\bibitem{li2021spatial}
G.~Li, C.-C. Hung, M.~Liu, L.~Pan, W.-C. Peng, and S.-H.~G. Chan,
  ``Spatial-temporal similarity for trajectories with location noise and
  sporadic sampling,'' in \emph{2021 IEEE 37th International Conference on Data
  Engineering (ICDE)}.\hskip 1em plus 0.5em minus 0.4em\relax IEEE, 2021.

\bibitem{kim2018scalable}
K.~S. Kim, S.~Lee, and K.~Huang, ``A scalable deep neural network architecture
  for multi-building and multi-floor indoor localization based on wi-fi
  fingerprinting,'' \emph{Big Data Anal.}, vol.~3, no.~1, pp. 1--17, 2018.

\bibitem{DeepWalk}
B.~Perozzi, R.~{Al-Rfou}, and S.~Skiena, ``{{DeepWalk}}: Online learning of
  social representations,'' in \emph{ACM KDD}, 2014.

\bibitem{Mikolov2013}
T.~Mikolov, I.~Sutskever, K.~Chen, G.~Corrado, and J.~Dean, ``Distributed
  representations of words and phrases and their compositionality,'' in
  \emph{NeurIPS}, 2013.

\bibitem{van2008visualizing}
L.~Van~der Maaten and G.~Hinton, ``Visualizing data using t-sne.'' \emph{J.
  Mach. Learn. Res.}, vol.~9, no.~11, 2008.

\bibitem{lee2013pseudo}
D.-H. Lee, ``Pseudo-label: The simple and efficient semi-supervised learning
  method for deep neural networks,'' in \emph{ICML Workshop}, 2013.

\bibitem{xie2020self}
Q.~Xie, M.-T. Luong, E.~Hovy, and Q.~V. Le, ``Self-training with noisy student
  improves imagenet classification,'' in \emph{IEEE CVPR}, 2020.

\bibitem{Goodfellow-et-al-2016}
I.~Goodfellow, Y.~Bengio, and A.~Courville, \emph{Deep Learning}.\hskip 1em
  plus 0.5em minus 0.4em\relax MIT Press, 2016.

\bibitem{cox2008multidimensional}
M.~A.~A. Cox and T.~F. Cox, ``Multidimensional scaling,'' in \emph{Handbook of
  Data Visualization}.\hskip 1em plus 0.5em minus 0.4em\relax Springer, 2008,
  pp. 315--347.

\bibitem{kdd05FeatureBagging}
A.~Lazarevic and V.~Kumar, ``Feature bagging for outlier detection,'' in
  \emph{ACM KDD}, 2005.

\bibitem{IsolationForest}
F.~T. Liu, K.~M. Ting, and Z.-H. Zhou, ``Isolation forest,'' in \emph{IEEE
  ICDM}, 2008.

\bibitem{lof}
M.~M. Breunig, H.-P. Kriegel, R.~T. Ng, and J.~Sander, ``{{LOF}}: Identifying
  density-based local outliers,'' in \emph{{{ACM SIGMOD}}}, 2000.

\bibitem{zweig1993receiver}
M.~H. Zweig and G.~Campbell, ``Receiver-operating characteristic (roc) plots: a
  fundamental evaluation tool in clinical medicine,'' \emph{Clin. Chem.}, 1993.

\bibitem{thomas2006elements}
M.~Thomas and A.~T. Joy, \emph{Elements of information theory}.\hskip 1em plus
  0.5em minus 0.4em\relax Wiley-Interscience, 2006.

\bibitem{torres2014ujiindoorloc}
J.~Torres-Sospedra, R.~Montoliu, A.~Mart{\'\i}nez-Us{\'o}, J.~P. Avariento,
  T.~J. Arnau, M.~Benedito-Bordonau, and J.~Huerta, ``{UJIIndoorLoc}: A new
  multi-building and multi-floor database for {WLAN} fingerprint-based indoor
  localization problems,'' in \emph{2014 International Conference on Indoor
  Positioning and Indoor Navigation (IPIN)}.\hskip 1em plus 0.5em minus
  0.4em\relax IEEE, 2014.

\end{thebibliography}
\bibliographystyle{IEEEtran}

\end{document}